\documentclass[a4paper,11pt]{article}
\pdfoutput=1 

\usepackage{jheppub} 
\usepackage{tikz}                     
\usetikzlibrary{shapes.geometric}
\usepackage{subfiles}
\usepackage{multirow}
\usepackage{subcaption}

\usepackage[T1]{fontenc} 

\newcommand{\iso}{\cong}

\newcommand{\iu}{{i\mkern1mu}}
\newcommand{\N}{\mathcal{N}}

\newcommand{\Z}{\mathbb{Z}}
\newcommand{\Zc}{\mathcal{Z}}

\newcommand{\inst}{\text{inst}}
\DeclareMathOperator{\tr}{tr}
\DeclareMathOperator{\Tr}{Tr}

\DeclareMathOperator{\diag}{diag}

\DeclareMathOperator{\PE}{\mathrm{PE}}

\DeclareMathOperator{\sign}{sign}

\title{\boldmath Instanton counting in Class $\mathcal{S}_k$ }

\preprint{DESY 17-189}

\author{Thomas Bourton,}
\author{Elli Pomoni}

\affiliation{DESY Theory Group, Notkestra{\ss}e 85, 22607 Hamburg, Germany}

\emailAdd{thomas.bourton@desy.de}
\emailAdd{elli.pomoni@desy.de}

\abstract{

\bigskip

We compute the instanton partition functions of $\mathcal{N}=1$ SCFTs in class $\mathcal{S}_k$. We obtain this result via orbifolding Dp/D(p-4) brane systems and calculating the partition function of the supersymmetric gauge theory on the worldvolume of $K$ D(p-4) branes.
Starting with D5/D1 setups probing a $\mathbb{Z}_\ell\times \mathbb{Z}_k$ orbifold singularity we obtain the $K$ instanton partition functions of 6d $(1,0)$ theories on $\mathbb{R}^4 \times T^2$ in the presence of orbifold defects on $T^2$ via computing the 2d superconformal index of the worldvolume theory on $K$ D1 branes wrapping the $T^2$. We then reduce our results to the 5d and to the 4d instanton partition functions. For $k=1$ we check that we reproduce the known elliptic, trigonometric and rational Nekrasov partition functions.
Finally, we show that the instanton partition functions of $SU(N)$ quivers in class $\mathcal{S}_k$  can be obtained from the class $\mathcal{S}$ mother theory partition functions with $SU(kN)$ gauge factors via imposing the `orbifold condition' $a_{\mathcal{A}} \rightarrow a_A e^{2\pi i j/k}$ with $\mathcal{A}=jA$ and $A=1,\dots, N$, $j=1,\dots, k$ on the Coulomb moduli and the mass parameters.
}

\begin{document} 
\maketitle
\flushbottom

\section{Introduction}

In recent years much progress has been made towards the non-perturbative study of four dimensional (4d) gauge theories with extended supersymmetry.
A milestone was the work of Seiberg and Witten who demonstrated  that the instanton series may be effectively summed by computing the periods of a holomorphic curve, known as the Seiberg-Witten (SW) curve \cite{Seiberg:1994aj,Seiberg:1994rs}. 
Nekrasov was able to verify their results from a purely field theoretic perspective and derive the instanton partition function by performing the integration over (a suitable regularization) of the instanton moduli space  \cite{Nekrasov:2003rj,Nekrasov:2002qd}.

String or M-theory realisations as well as compactifications of higher dimensional theories to 4d  have also shed  much light on  the structure of 4d $\N=2$ theories.
A large class of 4d $\N=2$ theories may be obtained via (twisted) compactifications of the 6d $(2,0)$ SCFT on  $\mathcal{M}_4\times\mathcal{C}$,
a 4d compact manifold $\mathcal{M}_4$ and
a Riemann surface $\mathcal{C}$.
The $\N=2$ theories obtained in this way are said to lie in class $\mathcal{S}$ \cite{Gaiotto:2009we,Gaiotto:2009hg}. 
What is more, many protected quantities, such as partition functions \cite{Pestun:2007rz} and correlation functions of BPS operators 
 in class $\mathcal{S}$ SCFTs may be computed as observables of a 2d theory which lives on $\mathcal{C}$ \cite{Alday:2009aq,Alday:2009fs}. One manifestation of this 4d/2d relation is that the partition function on an  ellipsoid $\mathcal{M}_4=\mathbb{S}^4_{\epsilon_1,\epsilon_2}$ is equal to correlators in Liouville/Toda CFT \cite{Alday:2009aq,Wyllard:2009hg}.  Moreover, the 2d Virasoro/W-algebra conformal blocks are mapped to Nekrasov's instanton partition function.
Another manifestation of a 4d/2d relation is the
 partition function on $\mathcal{M}_4=\mathbb{S}^3\times \mathbb{S}^1$ (a.k.a. the superconformal index: SCI), which can be recast as a correlator of a 2d TQFT living on $\mathcal{C}$ \cite{Gadde:2009kb,Gadde:2011ik}.

An  $\N=1$ offspring of the class $\mathcal{S}$ construction was recently proposed in \cite{Gaiotto:2015usa} and denoted as class $\mathcal{S}_k$
and further investigated in \cite{Bah:2017gph,Heckman:2016xdl,Razamat:2016dpl,Coman:2015bqq,Morrison:2016nrt,Franco:2015jna,Hanany:2015pfa}. See also \cite{Apruzzi:2016nfr,DelZotto:2017pti,Apruzzi:2017iqe,Hassler:2017arf}. These are 4d  $\N=1$ theories which may be obtained via (twisted) compactification of 6d $(1,0)$ SCFTs again on  $\mathcal{C}$. 
The landscape of 6d $(1,0)$ theories is far richer, a classification via F-theory has been explored \cite{Heckman:2013pva,DelZotto:2014hpa,Heckman:2015bfa,Bhardwaj:2015oru}, although, a complete field theoretic understanding is currently lacking. An interesting subset of 6d $(1,0)$ theories are the $\mathcal{T}^N_k$ theories which may be engineered within M-theory by considering the low energy theory living on $N$ coincident and parallel M$5$ branes at the tip of a transverse $\Gamma=A_{k-1}$ singularity. The construction may be generalised by choosing $\Gamma=ADE$  which, upon compactification on $\mathcal{C}$ leads to a bigger class of 4d  $\N=1$ theories
 denoted by class $\mathcal{S}_{\Gamma}$ \cite{Heckman:2016xdl}.

Because of their orbifold constructions class $\mathcal{S}_k$ theories may provide an ideal starting point to begin to look for exact results  and 2d/4d relations in 4d $\mathcal{N}=1$ theories which, so far, have been largely unexplored. Often the orbifolded daughter theory possesses many similarities with their mother theory   \cite{Bershadsky:1998mb,Bershadsky:1998cb}.

In \cite{Gaiotto:2015usa} the 4d SCI was computed and most strikingly it was recast as a correlator of a 2d TQFT establishing the first 2d/4d relation for class $\mathcal{S}_k$. Subsequently, in \cite{Ito:2016fpl,Maruyoshi:2016caf}, the index and its TQFT description was also computed in the presence of half-BPS surface defects. Additionally, in \cite{Coman:2015bqq}, SW curves were computed and some of their properties explored. In \cite{Mitev:2017jqj}, guided by the SW curves, the existence of an AGT-like correspondence for the class $\mathcal{S}_k$, denoted AGT$_k$ correspondence, was conjectured. Furthermore, the instanton partition function $\Zc_{\text{inst}}^{\mathcal{S}_k}$ was  proposed  via the relation of $\Zc_{\text{inst}}^{\mathcal{S}_k}$ and $\mathcal{W}_{kN}$ conformal blocks.

In this paper we will compute $\Zc^{\mathcal{S}_k}_{\text{inst}}$ and verify the proposal of \cite{Mitev:2017jqj} for a subset  of class $\mathcal{S}_k$ theories with a Lagrangian description; 4d theories obtained by compactification of $\mathcal{T}_{k}^N$ when the compactification surface $\mathcal{C}$ is either an $\ell$-punctured torus $\mathcal{C}_{1,\ell}=T^2\setminus\{p_1,\dots,p_\ell\}$ or an $\ell+2$-punctured sphere $\mathcal{C}_{0,\ell+2}=\mathbb{CP}^1\setminus\{p_1,\dots,p_{\ell+2}\}$. These theories are conformal and have weakly coupled Lagrangian descriptions in terms of toroidal  or cylindrical $\N=1$ quiver theories. 

We engineer the toroidal $\N=1$ quiver theories of class $\mathcal{S}_k$ in type IIB string theory as $\mathbb{Z}_\ell \times \mathbb{Z}_k$ orbifolds of D3 branes\footnote{This is a T-dual version of the type IIA  Hanany-Witten description \cite{Hanany:1996ie,Witten:1997sc} used in \cite{Gaiotto:2015usa}. 
}.
We can study the dynamics of $K$ instantons on the D$p$ branes via
the correspondence between the ADHM construction \cite{Atiyah:1978ri} and D$(p-4)$ branes within D$p$ branes \cite{Douglas:1996uz,Douglas:1995bn,Witten:1994tz,Tong:2005un,Polchinski:1996na,Dorey:2002ik,Dorey:2000zq}. 
Instantons of orbifold daughters of $\mathcal{N}=4$ SYM were intensively studied in the early days of AdS/CFT  \cite{Dorey:1999pd,Hollowood:1999bm,Billo:2002hm,Billo:2006jm,Billo:2007py,Argurio:2007vqa,Billo:2008sp} 
and recent computations in theories with eight or more supercharges in various dimensions  \cite{Hwang:2014uwa,Kim:2011mv,Kim:2017xan,Haghighat:2013tka,Haghighat:2013gba} have been possible due to
significant improvement of the old techniques.
For this paper we where especially inspired by \cite{Pan:2016fbl}.
  The instanton moduli space on  D$3$ branes is isomorphic to the Higgs branch of the theory on D$(-1)$ branes.
The partition function of the supersymmetric matrix model theory on the worldvolume of $K$ D$(-1)$ branes is equal to the $K$ instanton partition function of the corresponding  class $\mathcal{S}_k$ theory.

Using T-duality on the setup of D$3$ branes in the presence of a  $\mathbb{Z}_\ell \times \mathbb{Z}_k$ orbifold singularity we land on D$5$ branes in the presence of a $\mathbb{Z}_\ell \times \mathbb{Z}_k$ orbifold which engineers a 6d (elliptic) uplift of the 4d theories we are interested in.
The matrix model describing pointlike instantons of the 4d theory living on D3 branes is lifted to a 2d gauge theory, the SCI of which computes the instanton partition function of the corresponding 6d theory on $T^2$ \cite{Pan:2016fbl}, living on the worldvolume of the D5 branes. 
The 2d SCI calculation is very well studied \cite{Putrov:2015jpa,Gadde:2013ftv,Gadde:2014ppa,Gadde:2013lxa,Nakayama:2011pa,Cordova:2017ohl}. 
Taking the 4d limit of the 6d instanton partition function
we obtain the instanton partition function of the 4d class $\mathcal{S}_k$ theory.

This paper is organised as follows. In Section \ref{sec:stringsetup}, we present the string theory setup on which we base our calculations. Experts can skip this section, however we find it crucial for building up notation and intuition for the rest of the sections.
In Section \ref{sec:2Star_Instantons}, we prepare for our main calculation by practising with the calculation of the instanton partition function of 4d mass deformed $\mathcal{N}=4$ SYM (a.k.a. $\mathcal{N}=2^*$) and its 5d and 6d (trigonometric and elliptic) uplifts: mass deformed 5d $\mathcal{N}=2$ MSYM on $\mathbb{S}^1$ and  6d $(2,0)$ theory on $T^2$. This is obtained via the computation of the SCI of the $(4,4)$ 2d gauge theory living on the worldvolume of the $K$ D1 branes with quiver depicted in Figure \ref{fig:D1quiver}, which we set up using a supercharge that survives the orbifold projection that will come next. In Section \ref{sec:Orbifolding}, we present the main computation of our paper, we perform a $\mathbb{Z}_\ell \times \mathbb{Z}_k$ `orbifold' to the SCI of Section \ref{sec:2Star_Instantons}. 
We extract the instanton partition function and for
 $k=1$ successfully check our result against the known instanton partition function of $\mathcal{N}=2$ circular quivers as well as their 5d and 6d uplifts. 
 We conclude in Section \ref{sec:Conclusions} with a summary and a discussion of our findings as well as an outlook of future directions.
Technical details are presented in the appendix to not interrupt the flow of the main text.

\section{String theory description}
\label{sec:stringsetup}

In this section we present the brane setups which we use to `engineer' Lagrangian theories in class $\mathcal{S}_k$. We use this opportunity to establish notation and discuss the bosonic and fermionic symmetries of theories in class $\mathcal{S}_k$.
We begin with the toroidal $\N=1$ quiver theories in class $\mathcal{S}_k$  which are obtained using type IIB string theory with $N$ D3 branes probing a $\mathbb{Z}_\ell \times \mathbb{Z}_k$  orbifold singularity (Table \ref{table:D3D-1}).
They are examples of $\mathcal{N}=1$ orbifold daughters of $\mathcal{N}=4$ SYM \cite{Kachru:1998ys,Lawrence:1998ja} and were extensively studied in the early days of AdS/CFT. After T-duality we land on type IIA string theory with $N$ D4 and $\ell$ NS5 branes in the presence of a $\mathbb{Z}_k$ orbifold singularity (see Table \ref{table:D4D02}), which was used in  \cite{Gaiotto:2015usa},
and naturally produces cylindrical $\N=1$ quiver theories in class $\mathcal{S}_k$. 
Finally, we obtain a 6d uplift of the cylindrical quivers of class $\mathcal{S}_k$ after a further T-duality (see Table \ref{table:D4D02} and Table \ref{table:D5D1}) which leads to $N$ D5 branes on a $\mathbb{Z}_\ell \times \mathbb{Z}_k$ orbifold singularity. 

\subsection{Type IIB realisation}
\begin{figure}
\centering
  \begin{tikzpicture}[square/.style={regular polygon,regular polygon sides=4},thick,inner sep=0.1em,scale=0.9]
    \node (G11) at (0,0)[circle,draw,minimum size=1cm]{$N_{11}$};
    \node (G21) at (2,0) [circle,draw,minimum size=1cm]{$N_{21}$};
    \node (G31) at (4,0) [circle,draw,minimum size=1cm]{$N_{31}$};
    \node (G41) at (6,0) [circle,draw,minimum size=1cm]{$N_{41}$};
    \node (G51) at (8,0)[circle,draw,minimum size=1cm]{$N_{51}$};
    \node (G12) at (0,-2)[circle,draw,minimum size=1cm]{$N_{12}$};
    \node (G22) at (2,-2) [circle,draw,minimum size=1cm]{$N_{22}$};
    \node (G32) at (4,-2) [circle,draw,minimum size=1cm]{$N_{32}$};
    \node (G42) at (6,-2)[circle,draw,minimum size=1cm]{$N_{42}$};
    \node (G52) at (8,-2)[circle,draw,minimum size=1cm]{$N_{52}$};
	\node (G13) at (0,-4)[circle,draw,minimum size=1cm]{$N_{13}$};
    \node (G23) at (2,-4) [circle,draw,minimum size=1cm]{$N_{23}$};
    \node (G33) at (4,-4) [circle,draw,minimum size=1cm]{$N_{33}$};
    \node (G43) at (6,-4)[circle,draw,minimum size=1cm]{$N_{43}$};
    \node (G53) at (8,-4)[circle,draw,minimum size=1cm]{$N_{53}$};
    
    \draw [->,color=blue] (G11.270) to (G12.90);
    \draw [->,color=blue] (G21.270) to (G22.90);
    \draw [->,color=blue] (G31.270) to (G32.90);
    \draw [->,color=blue] (G41.270) to (G42.90);
    \draw [->,color=blue] (G51.270) to (G52.90);
    \draw [->,color=blue] (G12.270) to (G13.90);
    \draw [->,color=blue] (G22.270) to (G23.90);
    \draw [->,color=blue] (G32.270) to (G33.90);
    \draw [->,color=blue] (G42.270) to (G43.90);
    \draw [->,color=blue] (G52.270) to (G53.90);
    \draw [->,color=blue] (0,1) to (G11.90);
    \draw [->,color=blue] (2,1) to (G21.90);
    \draw [->,color=blue] (4,1) to (G31.90);
    \draw [->,color=blue] (6,1) to (G41.90);
    \draw [->,color=blue] (8,1) to (G51.90);
    \draw [-,color=blue] (G13.270) to (0,-5);
    \draw [-,color=blue] (G23.270) to (2,-5);
    \draw [-,color=blue] (G33.270) to (4,-5);
    \draw [-,color=blue] (G43.270) to (6,-5);
    \draw [-,color=blue] (G53.270) to (8,-5);
    
    \draw [->,color=green] (-1,0) to (G11.180);
    \draw [->,color=green](G11.0) to (G21.180);
    \draw [->,color=green] (G21.0) to (G31.180);
    \draw [->,color=green] (G31.0) to (G41.180);
    \draw [->,color=green] (G41.0) to (G51.180);
    \draw [-,color=green] (G51.0) to (9,0);
    \draw [->,color=green] (-1,-2) to (G12.180);
    \draw [->,color=green] (G12.0) to (G22.180);
    \draw [->,color=green] (G22.0) to (G32.180);
    \draw [->,color=green] (G32.0) to (G42.180);
    \draw [->,color=green] (G42.0) to (G52.180);
    \draw [-,color=green] (G52.0) to (9,-2);
    \draw [->,color=green] (-1,-4) to (G13.180);
    \draw [->,color=green] (G13.0) to (G23.180);
    \draw [->,color=green] (G23.0) to (G33.180);
    \draw [->,color=green] (G33.0) to (G43.180);
    \draw [->,color=green] (G43.0) to (G53.180);
    \draw [-,color=green] (G53.0) to (9,-4);

    \draw [-,color=red] (G11.130) to (-1,1);
    \draw [-,color=red] (G21.130) to (1,1);
    \draw [-,color=red] (G31.130) to (3,1);
    \draw [-,color=red] (G41.130) to (5,1);
    \draw [-,color=red] (G51.130) to (7,1);
    \draw [-,color=red] (G12.130) to (-1,-1);
    \draw [-,color=red] (G13.130) to (-1,-3);

	\draw [<-,color=red] (G11.-40) to (G22.130);
    \draw [<-,color=red] (G21.-40) to (G32.130);
    \draw [<-,color=red] (G31.-40) to (G42.130);
    \draw [<-,color=red] (G41.-40) to (G52.130);
    \draw [<-,color=red] (G12.-40) to (G23.130);
    \draw [<-,color=red] (G22.-40) to (G33.130);
    \draw [<-,color=red] (G32.-40) to (G43.130);
    \draw [<-,color=red] (G42.-40) to (G53.130);
    
    \draw [<-,color=red] (G13.-40) to (1,-5);
    \draw [<-,color=red] (G23.-40) to (3,-5);
    \draw [<-,color=red] (G33.-40) to (5,-5);
    \draw [<-,color=red] (G43.-40) to (7,-5);
    \draw [<-,color=red] (G53.-40) to (9,-5);
    \draw [<-,color=red] (G52.-40) to (9,-3);
    \draw [<-,color=red] (G51.-40) to (9,-1);
    
    \draw [->] (-1,1) to (4,1);
    \draw [-] (4,1) to (9,1);
    \draw [->] (-1,-5) to (4,-5);
    \draw [-] (4,-5) to (9,-5);
    \draw [->>] (-1,-5) to (-1,-2);
    \draw [-] (-1,-2) to (-1,1);
    \draw [->>] (9,-5) to (9,-2);
    \draw [-] (9,-2) to (9,1);
  \end{tikzpicture}
  \caption{\it The quiver diagram in $\N=1$ notation with $\ell=5$, $k=3$. Circular nodes denote vector multiplets and coloured arrows denote chiral multiplets. Blue lines denote $\Phi^1_{(n,i)}$, green $\Phi^2_{(n,i)}$ and red $\Phi^3_{(n,i)}$.  The quiver should be periodically identified in both `$\ell$' and `$k$' directions, with gluing indicated by the black arrowed lines, such that it has the topology of a torus.}
  \label{fig:Skquiver}
\end{figure}
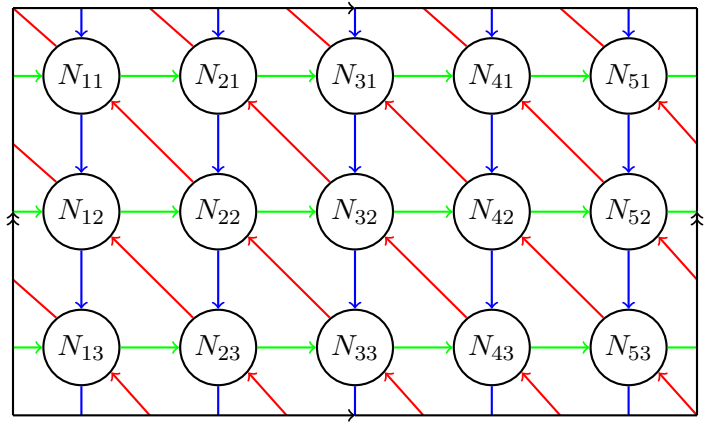

Consider Type IIB string theory on $\mathbb{R}^4\times\mathbb{R}^6/\Gamma$ with $\Gamma=\mathbb{Z}_{\ell}\times\mathbb{Z}_k$ with $\ell,k\in\mathbb{Z}^+$. Our goal is to engineer a certain subset of class $\mathcal{S}_k$ theories within Type IIB string theory. Hence, we add a set of $N$ parallel and coincident D$3$ branes along the $\mathbb{R}^4$ as described in Table \ref{table:D3D-1}. We parametrise the worldvolume of the D$3$ branes with four real coordinates $X^1,X^2,X^3,X^4$, which arrange themselves into the vector representation of $Spin(4)\iso SU(2)_{\alpha}\times SU(2)_{\dot\alpha}$. The Cartans, $J_L,J_R$, of $\mathfrak{su}(2)_{\alpha},\mathfrak{su}(2)_{\dot\alpha}$ are defined such that lower $\alpha=1,2$ have $J_L=+\frac{1}{2},-\frac{1}{2}$ and $\dot{\alpha}=\dot1,\dot2$ have $J_R=+\frac{1}{2},-\frac{1}{2}$. The $\mathbb{R}^6\iso\mathbb{C}^3$ is parametrised by six real coordinates $X^5,X^6,X^7,X^8,X^9,X^{10}$ and the isomorphism is made by the choice of arrangement into the complex coordinates
\begin{equation}
\label{eqn:coordinates}
Z_{56}
:=\frac{X^5+\iu X^6}{\sqrt{2}}=\Phi^1|_{\theta=0},\quad 
Z_{710}
:=\frac{X^7+\iu X^{10}}{\sqrt{2}}=\Phi^2|_{\theta=0},\quad 
Z_{89}
:=\frac{X^8+\iu X^9}{\sqrt{2}}=\Phi^3|_{\theta=0}
\end{equation} 
and their hermitian conjugates.
\begin{table}[ht]
\centering
\begin{tabular}{ |c |c| c| c| c| c| c| c| c| c| c| }
\hline
   & $X^1$ & $X^2$ & $X^3$ & $X^4$ & $X^5$ & $X^6$ & $X^7$ & $X^8$ & $X^9$& $X^{10}$\\\hline 
 $N$ D$3$ & -- & -- & -- & -- & $\cdot$ & $\cdot$ & $\cdot$ & $\cdot$ & $\cdot$ & $\cdot$\\ \hline
 $A_{\ell-1}$ & $\cdot$ & $\cdot$ & $\cdot$ & $\cdot$ & $\cdot$ & $\cdot$ & $\times$ & $\times$ & $\times$ & $\times$\\ \hline
 $A_{k-1}$ & $\cdot$ & $\cdot$ & $\cdot$ & $\cdot$ & $\times$ & $\times$ & $\cdot$ & $\times$ & $\times$ & $\cdot$\\ \hline
 \hline
 $K$ D$(-1)$ & $\cdot$ & $\cdot$ & $\cdot$ & $\cdot$ & $\cdot$ & $\cdot$ & $\cdot$ & $\cdot$ & $\cdot$ & $\cdot$ \\\hline
\end{tabular}
\caption{\it The type IIB setup engineering Lagrangian 4d SCFTs  in class $\mathcal{S}_k$.}
\label{table:D3D-1}
\end{table}
The orbifold $\Gamma$ acts on those coordinates \eqref{eqn:coordinates} as
\begin{equation}\label{eqn:orbifoldactionR10}
\Gamma:\left(Z_{56} , Z_{710}  ,Z_{89} \right)\mapsto \left(\omega_k Z_{56} ,\omega_{\ell} Z_{710} ,\omega_{\ell}^{-1}\omega_{k}^{-1}Z_{89} \right)
\end{equation}
where $\omega_{\ell}:=e^{2\pi\iu/\ell}$, $\omega_{k}:=e^{2\pi\iu/k}$. Before the orbifold action, fundamental strings stretching between D$3$ branes give rise to the $SU(N)$ $\N=4$ SYM theory on their worldvolume; with $R$-symmetry $Spin(6)_R\iso SU(4)_R$, the rotation group of the transverse $\mathbb{R}^6$ spanned by $X^5,X^6,X^7,X^8,X^9,X^{10}$. In $\N=1$ superspace the theory contains a vector multiplet $\mathcal{V}$ and three chiral superfields in the adjoint of the gauge group: $\left(\Phi^1,\Phi^2,\Phi^3\right)^T$ transforming in the $\mathbf{3}$ of $SU(3)_R\subset SU(4)_R$. The superpotential is given by
\begin{equation}
\mathcal{W}_{\N=4}=\frac{\iu}{3!}\epsilon_{abc}\tr\Phi^a\left[\Phi^b,\Phi^c\right]-\iu\frac{\tau_{YM}}{8\pi}\tr W^{\alpha}W_{\alpha}\,.
\end{equation}
The chiral superfields $\Phi^a$ are identified with transverse coordinates \eqref{eqn:coordinates} hence the action of $\Gamma$ on $\mathbb{C}^3$ lies diagonally inside $SU(3)_{R}$ in the form
\begin{equation}
M_a^b:=\begin{pmatrix}
\omega_k&0&0\\
0&\omega_{\ell}&0\\
0&0&\omega_{\ell}^{-1}\omega_{k}^{-1}
\end{pmatrix}^b_a\in SU(3)_{R}\,.
\end{equation}
Note that $\Gamma$ also has an action inside the gauge group, $SU(N)$ \cite{Douglas:1996sw}. Its action can be conjugated to an element $h$ of the maximal torus $T(SU(N))=U(1)^{N-1}$. After scaling $N\to|\Gamma|N=\ell kN$ this action breaks $G\to \prod_{n=1}^{\ell}\prod_{i=1}^kSU(N_{ni})$ specified by an $\ell k$ partition of $\ell kN=\sum_{n,i}N_{ni}$. Note we always take the orbifold indices to be $n,m=1,\dots,\ell$ and $i,j=1,\dots,k$ and we impose orbifold periodicity $n\sim n+\ell$, $i\sim i+k$. There is a unique way to preserve conformal invariance; we choose the action of $\Gamma$ such that $N_{ni}=N$ for all $n,i$. Hence $h$ may be written as
\begin{equation}\label{eqn:orbifoldactionSUN}
h=\diag\left(\omega_{\ell}\omega_k\mathbb{I},\dots,\omega_{\ell}\omega_k^k\mathbb{I},\dots,\omega^{\ell}_{\ell}\omega_k\mathbb{I},\dots, \omega_{\ell}^{\ell}\omega^k_k\mathbb{I}\right)
\end{equation}
where $\mathbb{I}$ denotes the $N\times N$ identity matrix. Quotienting by $\Gamma$ imposes the identifications
\begin{equation}
\mathcal{V}\sim h^{\dagger}\mathcal{V}h\,,\quad \Phi^a\sim M^a_bh^{\dagger}\Phi^bh\,.
\end{equation}
After performing these identifications the resulting theory is an $\N=1$ torodial quiver gauge theory with gauge group $SU\left(N\right)^{\ell k}$ and superpotential
\begin{equation}
\mathcal{W}_{\N=1}=\sum_{n=1}^{\ell}\sum_{i=1}^k\iu\Phi_{(n,i)}^1\left(\Phi_{(n-1,i)}^2\Phi_{(n,i-1)}^3-\Phi_{(n+1,i+1)}^2\Phi_{(n,i+1)}^3\right)-\iu\frac{\tau_{YM,ni }}{8\pi}\tr W^{\alpha}_{ni}W_{ni,\alpha}
\end{equation} 
which now transform as $\Phi_{(n,i)}^1\in (N_{ni},\overline{N}_{n(i+1)})$, $\Phi_{(n,i)}^2\in (N_{ni},\overline{N}_{(n+1)i})$ and $\Phi_{(n,i)}^3\in (N_{ni},\overline{N}_{(n-1)(i+1)})$ under the gauge group $\prod_{n,i}SU(N_{ni})=SU\left(N\right)^{\ell k}$. We summarise the field content in the quiver diagram of Figure \ref{fig:Skquiver}. The individual couplings for each gauge node $g_{YM,ni}^2$ are given by integration of a non-zero $B$-field flux over the two-cycles $C_{ni}$ of the space obtained by resolving the $\mathbb{C}^3/\Gamma$ singularities
\begin{equation}\label{eqn:Bfield}
\int_{C_{ni}}B=\frac{4\pi^2}{g^2_{YM,ni}}\,,\quad \sum_{n,i}\frac{1}{g_{YM,ni}^2}=\frac{1}{g^2_{YM}}\,.
\end{equation}
These are precisely the same class of $\N=1$ SCFTs which we expect to describe, at low energies, the 4d thoery obtained by placing $N$ M$5$ branes at the tip of an $A_{k-1}$ singularity (which is the family of the 6d $\N=(1,0)$ $\mathcal{T}^N_k$ theories), compactified on $T^2$ with $\ell$ punctures and complex structure $\tau_{YM}=\frac{4\pi\iu}{g_{YM}^2}+  \frac{\theta}{2\pi}$ \cite{Gaiotto:2015usa}.  This statement is known to be true \cite{Gaiotto:2015usa} at the orbifold point for the $\mathbb{Z}_k$ orbifold, which means that all $g_{YM,ni}^2= g_{YM,n}^2$ for all $i=1,\dots,k$. There, the $\ell$ different coupling constants $g_{YM,n}^2$ correspond to the position of the  $\ell$ punctures of the torus. What happens away from the  orbifold point for the $\mathbb{Z}_k$ orbifold is currently under investigation \cite{Tom}.

\subsection{Type IIA realisation}
Indeed, by performing a T-duality along, say, $X^7$ to the setup of Table \ref{table:D3D-1} we may obtain the Hanany-Witten description of the above class $\mathcal{S}_k$ theories in Type IIA as described in \cite{Gaiotto:2015usa}. To perform the T-duality we may partially resolve the $\mathbb{C}^3/\Gamma$ singularity. Resolving the $A_{\ell-1}$ singularity gives rise to an ALE space which is equivalent to the $\lambda\to\infty$ limit of the $\ell$-centred Taub-Nut space TN$_{\ell}$ with metric
\begin{equation}
ds^2=V^{-1}\left(d\Theta+\vec{A}\cdot d\vec{x}\right)^2+Vd\vec{x}^2,\quad V=\sum_{n=1}^{\ell}\frac{1}{\left|\vec{x}-\vec{x}_n\right|}+\frac{1}{\lambda^2}
\end{equation}
subject to the condition $\vec{\nabla}V=-\vec{\nabla}\times\vec{A}$ . The underlying geometry is that of an $\mathbb{S}^1$ fibered over an $\mathbb{R}^3$ base. To perform the T-duality we hence replace $\mathbb{C}^3/\Gamma$ by $\left(\mathbb{C}\times \text{TN}_{\ell}\right)/\Z_{k}$ where $\mathbb{C}$ is parametrised by $Z_{56},\overline{Z}_{56}$ and TN$_{\ell}$ is parametrised by $\Theta=X^7$, $\vec{x}=\left(X^8,X^9,X^{10}\right)$. We may then T-dualise along the TN$_{\ell}$ circle (which is invariant under the $\Z_{k}$ action). We hence obtain the Hanany-Witten description shown in Table \ref{table:D4D01} .
\begin{table}[h!]
\centering
\begin{tabular}{ |c |c| c| c| c| c| c| c| c| c| c| }
\hline
   & $X^1$ & $X^2$ & $X^3$ & $X^4$ & $X^5$ & $X^6$ & $X^7$ & $X^8$ & $X^9$& $X^{10}$\\\hline 
 $N$ D$4$ & -- & -- & -- & -- & $\cdot$ & $\cdot$ & -- & $\cdot$ & $\cdot$ & $\cdot$\\ \hline
 $\ell$ NS$5$ & -- & -- & -- & -- & -- & -- & $\cdot$ & $\cdot$& $\cdot$ & $\cdot$\\ \hline
 $A_{k-1}$ & $\cdot$ & $\cdot$ & $\cdot$ & $\cdot$ & $\times$ & $\times$ & $\cdot$ & $\times$ & $\times$ & $\cdot$\\ \hline
 \hline
 $K$ D$0$ & $\cdot$ & $\cdot$ & $\cdot$ & $\cdot$ &$\cdot$ & $\cdot$ & -- & $\cdot$ & $\cdot$ & $\cdot$ \\\hline
\end{tabular}
\caption{\it The type IIA setup obtained by a T-duality along $X^7$ to Table \ref{table:D3D-1}.}
\label{table:D4D01}
\end{table}
Under the T-duality the $\ell$ centers of TN$_{\ell}$ become $\ell$ NS$5$ branes fixed at positions $\Theta_n$ and $\vec{x}_n$ in the transverse directions. The angles $\Theta_n$ at which the NS$5$ branes sit along the $\mathbb{S}^1$ with radius $\beta_7$ are related to the gauge couplings \label{eqn:Bfield} at the $n$\textsuperscript{th} node by 
\begin{equation}
\frac{4\pi^2}{g_{YM,ni}}=\frac{\Theta_n-\sum_{a=1}^{n-1}\Theta_a}{(1-k)2\pi \beta_7}\int_{C_{i}}B,\quad \beta_7=\frac{1}{\lambda}
\end{equation}
with $C_{j}$ the two cycles of the resolution of $A_{k-1}$.

We could just as well have instead resolved the $A_{k-1}$ singularities and T-dualised along $X^5$ yielding the brane setup of Table \ref{table:D4D02}.
\begin{table}[h!]
\centering
\begin{tabular}{ |c |c| c| c| c| c| c| c| c| c| c| }
\hline
   & $X^1$ & $X^2$ & $X^3$ & $X^4$ & $X^5$ & $X^6$ & $X^7$ & $X^8$ & $X^9$& $X^{10}$\\\hline 
 $N$ D$4$ & -- & -- & -- & -- & -- & $\cdot$ & $\cdot$ & $\cdot$ & $\cdot$ & $\cdot$\\ \hline
 $A_{\ell-1}$ & $\cdot$ & $\cdot$ & $\cdot$ & $\cdot$ & $\cdot$ & $\cdot$ & $\times$ & $\times$& $\times$ & $\times$\\ \hline
 $k$ NS$5$ & -- & -- & -- & -- & $\cdot$ & $\cdot$ & -- & $\cdot$ & $\cdot$ & --\\ \hline
 \hline
 $K$ D$0$ & $\cdot$ & $\cdot$ & $\cdot$ & $\cdot$ & -- & $\cdot$ & $\cdot$ & $\cdot$ & $\cdot$ & $\cdot$ \\\hline
\end{tabular}
\caption{\it Alternative type IIA setup obtained by instead a T-duality along $X^5$ to Table \ref{table:D3D-1}.}
\label{table:D4D02}
\end{table}
Let us conclude this section with a final comment. Naively we see that there is a $k\leftrightarrow\ell$ symmetry. However, this is true only at the `orbifold point' for both orbifolds or for very special values of $k,\ell$ and the gauge couplings. The $A_{k-1}$ singularity has $k-1$ blow up modes, while $A_{\ell-1}$ has $\ell-1$. In case we turn them all on to generic points the $k\leftrightarrow\ell$ symmetry is lost.

\subsection{A 6d uplift}
\begin{table}[h!]
\centering
\begin{tabular}{ |c |c| c| c| c| c| c| c| c| c| c| }
\hline
   & $X^1$ & $X^2$ & $X^3$ & $X^4$ & $X^5$ & $X^6$ & $X^7$ & $X^8$ & $X^9$& $X^{10}$\\\hline 
 $N$ D$5$ & -- & -- & -- & -- & -- & -- & $\cdot$ & $\cdot$ & $\cdot$ & $\cdot$\\ \hline
 $A_{\ell-1}$ & $\cdot$ & $\cdot$ & $\cdot$ & $\cdot$ & $\cdot$ & $\cdot$ & $\times$ &$\times$ & $\times$ & $\times$\\ \hline
 $A_{k-1}$ & $\cdot$ & $\cdot$ & $\cdot$ & $\cdot$ & $\times$ & $\times$ & $\cdot$ & $\times$ & $\times$ & $\cdot$\\ \hline
 \hline
 $K$ D$1$ & $\cdot$ & $\cdot$ & $\cdot$ & $\cdot$ & -- & -- & $\cdot$ & $\cdot$ & $\cdot$ & $\cdot$ \\\hline
\end{tabular}
\caption{\it Type IIB setup engineering a 6d uplift of the 4d theories we are interested in.}
\label{table:D5D1}
\end{table}

In this paper our primary interest will be that of the theory living on the D$(-1)$/D$0$ branes in Tables \ref{table:D3D-1} , \ref{table:D4D01} and \ref{table:D4D02} . In both cases these are supersymmetric matrix models invariant under at least two supercharges. We could choose to work directly with these matrix models, however we find it more convenient to work instead with the two dimensional uplift of those matrix models. Hence, we instead work with the brane setup of Table \ref{table:D5D1}, obtained by performing a further T-duality along $X^6$ to Table \ref{table:D4D02}. Before performing the T-duality we also assume that $X^5$ may be safely decompactified such that it parametrises a space with the topology of $\mathbb{R}$. The T-duality may again be performed by replacing $A_{\ell-1}$ with TN$_{\ell}$ , we then T-dualise along the TN$_{\ell}$ circle, landing us on the setup of Table \ref{table:D5D1} .
The $Spin(6)_{R}$ $R$-symmetry group has been broken to a subgroup $U(1)_{56}\times Spin(4)_R$ which acts by rotations along $\mathbb{R}^2$ , $\mathbb{R}^4$ parametrised by $X^5,X^6$ and $X^7,X^8,X^9,X^{10}$ respectively. The $\mathbb{Z}_{\ell}$ orbifold further breaks the $Spin(4)_R$ $R$-symmetry group down to a subgroup $SU(2)_{R}$ corresponding to the isometry group of TN$_{\ell}$, while, the $\mathbb{Z}_k$ orbifold breaks the $U(1)_{56}\times SU(2)_R$ down to the maximal torus of $SU(2)_R$.

\begin{figure}[h]\label{fig:sub}
\begin{subfigure}[c]{0.5\linewidth}
\centering
\begin{tikzpicture}[thick,scale=1.2]
\node (A1) at (-2,4) {6d $(2,0)$ $A_{N-1}$}; 
\node [align=center] (A2) at (-2,2) {5d $SU(N)$ \\ $\mathcal{N}=2$ MSYM}; 
\node (B1) at (2,4) {6d $(1,0)$ $\mathcal{T}^N_{\ell}$}; 
\node (B2) at (2,2) {5d $\mathcal{N}_{N,\ell}$}; 
\node (C) at (0,0) {4d class $\mathcal{S}$ $\mathcal{N}=2$ $\tilde{A}_{\ell-1}$ }; 

\draw[->] (A1) -- (A2) node[midway,left] {$\mathbb{S}^1_6$};
\draw[->] (B1) -- (B2) node[midway,right] {$\mathbb{S}^1_6$};
\draw[->] (B2) -- (C) node[midway,right] {$\mathbb{S}^1_5$};
\draw[->] (A2) -- (C) node[midway,left,align=center] {$\mathbb{S}^1_5$\\Nahm pole BCs};
\end{tikzpicture}
\caption{\label{ComplactificationsWithZell}}
\end{subfigure}
\hspace*{\fill}
\begin{subfigure}[c]{0.5\linewidth}
\centering
\begin{tikzpicture}[thick, scale=1.2]
\node (A1) at (-2,4) {6d $(1,0)$ $\mathcal{T}^N_k$}; 
\node (A2) at (-2,2) {5d  $\mathcal{N}_{N,k}$}; 
\node[align=center] (B1) at (2,4) {6d $(1,0)$ $\mathcal{T}^{N}_{\ell}$\\ + defect}; 
\node[align=center] (B2) at (2,2) {5d $\mathcal{N}_{ N,\ell}$\\ + defect}; 
\node (C) at (0,0) {4d class $\mathcal{S}_k$ $\mathcal{N}=1$ $\tilde{A}_{\ell-1}\times\tilde{A}_{k-1}$ }; 

\draw[->] (A1) -- (A2) node[midway,left] {$\mathbb{S}^1_6$};
\draw[->] (B1) -- (B2) node[midway,right] {$\mathbb{S}^1_6$};
\draw[->] (B2) -- (C) node[midway,right] {$\mathbb{S}^1_5$};
\draw[->] (A2) -- (C) node[midway,left,align=center] {$\mathbb{S}^1_5$\\`Orbifold'- \\Nahm pole BCs};
\end{tikzpicture}
\caption{\label{ComplactificationsWithZellZk}}
\end{subfigure}
\caption{Left: \it Schematic overview for $k=1$ of two alternate ways to obtain the 4d $\N=2$ $\tilde{A}_{\ell-1}$ circular quivers with $SU(N)^\ell$ gauge group in class $\mathcal{S}$ from compactifications of 6d SCFTs. \normalfont Right: \it A schematic overview of the $k>1$ generalisations of compactifications of 6d SCFTs. The resulting 4d SCFTs are $\N=1$ $\tilde{A}_{\ell-1}\times \tilde{A}_{k-1}$ torodial quivers in class $\mathcal{S}_k$ with gauge group $SU(N)^{\ell k}$.}
\end{figure}
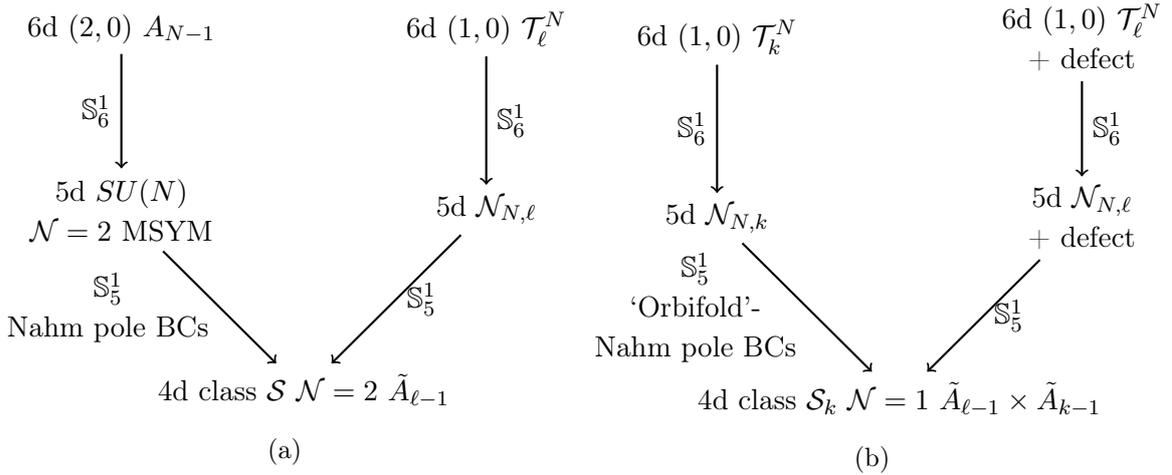
It is also useful to understand this IIB setup via uplifting it to M-theory - see Table \ref{table:M5M2}.
\begin{table}[h!]
\centering
\begin{tabular}{ |c |c| c| c| c| c| c| c| c| c| c| c| }
\hline
   & $X^1$ & $X^2$ & $X^3$ & $X^4$ & $X^5$ & $X^6$ & $X^7$ & $X^8$ & $X^9$& $X^{10}$ & $X^{11}$ \\\hline 
 $N$ M$5$ & -- & -- & -- & -- & -- & -- & $\cdot$ & $\cdot$ & $\cdot$ & $\cdot$ & $\cdot$ \\ \hline
 $A_{\ell-1}$ & $\cdot$ & $\cdot$ & $\cdot$ & $\cdot$ & $\cdot$ & $\cdot$ & $\times$ &$\times$ & $\times$ & $\times$ & $\cdot$ \\ \hline
 $A_{k-1}$ & $\cdot$ & $\cdot$ & $\cdot$ & $\cdot$ & $\times$ & $\times$ & $\cdot$ & $\times$ & $\times$ & $\cdot$ & $\cdot$ \\ \hline
 \hline
 $K$ M$2$ & $\cdot$ & $\cdot$ & $\cdot$ & $\cdot$ & -- & -- & $\cdot$ & $\cdot$ & $\cdot$ & $\cdot$ & --  \\\hline
\end{tabular}
\caption{\it The M-theory uplift of the IIB setup in table \ref{table:D5D1}, `engineering' a 6d uplift of the 4d theories we are interested in. For $k=1$ this is the same as (2.2) in \cite{Haghighat:2013tka}.
For  $\ell=1$, this is the 
codimension-two defect
 of Kanno and Tachikawa \cite{Kanno:2011fw}. }
\label{table:M5M2}
\end{table}  
We begin with the more familiar $k=1$ case, the 6d $(1,0)$ $\mathcal{T}^N_{\ell}$ SCFT associated to $N$ M5 branes sitting at the tip of an $\mathbb{Z}_{\ell}$ orbifold singularity of M-theory.
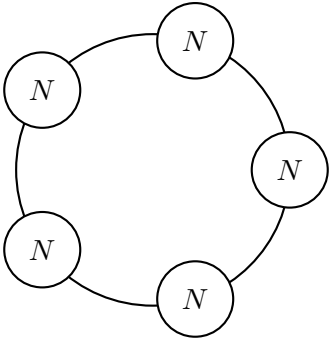
\begin{figure}
\centering
\begin{tikzpicture}[square/.style={regular polygon,regular polygon sides=4},thick,inner sep=0.1em]
   \newdimen\R
   \R=1.8cm
   \draw (0,0) circle (\R);
   \foreach \x/\l/\p in
     { 72/{3},
      144/{2},
      216/{1},
      288/{6},
      360/{4}
     }
     \node[circle,draw,minimum size=1.0cm,fill=white] at (\x:\R) {$N$};
\end{tikzpicture}
\caption{\it 5d circular (necklace) quiver $\mathcal{N}_{N,\ell}$ for $\ell=5$. Circular nodes denote $\mathcal{N}=1$ vector multiplets and solid lines connecting them denote bifundamental $\mathcal{N}=1$ hypermultiplets. Circle reduction of $\mathcal{N}_{N,\ell}$ results in 4d $\mathcal{N}=2$ circular $\tilde{A}_{\ell-1}$ quiver with the same structure.}
\label{fig:5dNNl}
\end{figure}
  Compactifying that on $\mathbb{S}^1_6$ gives 5d $\N=1$ circular quivers $\mathcal{N}_{N,\ell}$ with $\ell$ nodes denoting $SU(N)$ gauge groups\footnote{We wish to remind the reader that Fiber-Base duality exchanges $N\leftrightarrow\ell$ and leads to a duality between the 5d $\N=1$ circular quivers $\mathcal{N}_{N,\ell}$ and $\mathcal{N}_{\ell, N}$ \cite{Katz:1997eq,Bao:2011rc,Hohenegger:2013ala,Mitev:2014jza,DelZotto:2015rca,Bastian:2017ary}.} and $\ell$ links denoting bifundamental hypermultiplets \cite{Ohmori:2015pua,Ohmori:2015pia,Bah:2017gph}, see Figure \ref{fig:5dNNl}. Further compactification on $\mathbb{S}^1_5$ results in 4d $\N=2$ $\tilde{A}_{\ell-1}$ circular quiver theories with $SU(N)^{\ell}$ gauge group. The $\tilde{A}_{\ell-1}$ theory may also be realised via the well known Class $\mathcal{S}$ construction obtained by compactifying the $A_{N-1}$ $(2,0)$ theory on the $\ell$ punctured torus with certain half-BPS Nahm pole boundary conditions specified at the punctures \cite{Chacaltana:2012zy,Tsimpis:1998zh,Xie:2013gma}, see Figure \ref{ComplactificationsWithZell}.
  
When $k>1$ the resulting 6d theory corresponds to $\mathcal{T}^{N}_{\ell}$ in the presence of 
a codimension-two Gukov-Witten \cite{Gukov:2006jk,Gukov:2008sn} surface operator 
 associated to $\ell$ copies of the partition 
\begin{equation}\label{eqn:levipartition}
kN=N_1+\dots+N_k=N+\dots+N
\end{equation}
for the factors of $SU\left(kN\right)^{\ell}$. 
For  $\ell=1$, this is the 
codimension-two defect
 of Kanno and Tachikawa \cite{Kanno:2011fw} which may be realised as a $\mathbb{Z}_k$ orbifold in M-theory, see Table \ref{table:M5M2}. Finally, KK reducing along the circle on which we performed the final T-duality (from Table \ref{table:D4D02} to Table \ref{table:D5D1}) leads to the 5d $\N=1$ circular/necklace quiver gauge theory $\mathcal{N}_{kN,\ell}$ on $\mathbb{R}^4\times \mathbb{S}^1$ in the presence of the defect along the circle (see Figure \ref{ComplactificationsWithZellZk}). Analogously, there is also the Class $\mathcal{S}_k$ construction obtained by compactification of the $\mathcal{T}^N_k$ on a torus with $\ell$ punctures with `orbifold' Nahm pole boundary conditions specified at each puncture \cite{Gaiotto:2015usa,Heckman:2016xdl,Hassler:2017arf}.

\begin{figure}[t]

\end{figure}

\subsection{On supersymmetry of the D1/D5 system}
Type IIB string theory has $32$ supersymmetries parametrised by two $32$ component spinors $\epsilon_L$ , $\epsilon_R$ of positive chirality $\Gamma^{11}\mathcal{\epsilon}_{L/R}=+\mathcal{\epsilon}_{L/R}$ where $\Gamma^{11}=\Gamma^1\dots\Gamma^{10}$ and $\Gamma^M$ are the $32\times32$ Gamma matrices.
The D$5$/D$1$ system preserves $1/4$ of the $32$ supersymmetries. Between them they preserve only those supersymmetries of the form
\begin{equation}
\epsilon_L=\Gamma^1\Gamma^2\Gamma^3\Gamma^4\Gamma^5\Gamma^6\epsilon_R,\quad \epsilon_L=\sigma\Gamma^5\Gamma^6\epsilon_R,
\end{equation} 
with $\sigma=\pm1$ corresponding to whether we choose to insert D$1$ or anti-D$1$ branes. The theory living on the (anti-)D$1$ branes then possesses $(p,q)$ supersymmetry with $p+q=32/4=8$. By choosing an explicit representation for the Gamma matrices it can be shown that $p=q=4$ and that the preserved supercharges are 
\begin{align}
&Q^{\alpha a}_{+\frac{1}{2}},\quad Q^{\alpha \dot a}_{-\frac{1}{2}},\quad \text{if $\sigma=+1$}\\
&\overline{Q}^{\dot\alpha\dot a}_{+\frac{1}{2}},\quad\overline{Q}^{\dot\alpha a}_{-\frac{1}{2}},\quad \text{if $\sigma=-1$}
\end{align}
where $a,\dot a=1,2$ are indices of $Spin(4)\iso SU(2)_{a}\times SU(2)_{\dot a}$ and the subscript $\pm\frac{1}{2}$ denotes the representation under the $U(1)_{56}$ which acts as the Lorentz group of the D$1$-brane worldvolume theory.

The $SU(2)_{a}\times SU(2)_{\dot a}$ rotates the two planes of the $\mathbb{C}^2$ parametrised by $Z_{710},Z_{89}$ into one another. The Cartans of $\mathfrak{su}(2)_{a},\mathfrak{su}(2)_{\dot a}$ $J^R_{L},J_R^R$ may be expressed in terms of the generators $J_{710}$ and $J_{89}$ of $U(1)$ rotations in their respective planes as
\begin{equation}
\label{eq:RsymmetryRotations}
J_L^R=\frac{1}{2}\left(J_{710}-J_{89}\right),\quad J_R^R=\frac{1}{2}\left(J_{710}+J_{89}\right),
\end{equation}
which are defined such that lower $a=1,2$ have $J_L^R=+\frac{1}{2},-\frac{1}{2}$ and lower $\dot{a}=\dot1,\dot2$ have $J_R^R=+\frac{1}{2},-\frac{1}{2}$.  Hence the $\Gamma$ action on the supercharges is 
\begin{align}
\label{Qorbifold1}
&\Gamma:\left(Q^{\alpha a}_{+\frac{1}{2}},Q^{\alpha \dot a}_{-\frac{1}{2}}\right)\mapsto\omega_{\ell}^{2J_L^R}\omega_k^{J_{56}+J_L^R-J_R^R}\left(Q^{\alpha a}_{+\frac{1}{2}},Q^{\alpha \dot a}_{-\frac{1}{2}}\right)\,,\\\
\label{Qorbifold2}
&\Gamma:\left(\overline{Q}^{\dot\alpha\dot a}_{+\frac{1}{2}},\overline{Q}^{\dot\alpha a}_{-\frac{1}{2}}\right)\mapsto\omega_{\ell}^{2J_R^L}\omega_k^{J_{56}+J_L^R-J_R^R}\left(\overline{Q}^{\dot\alpha\dot a}_{+\frac{1}{2}},\overline{Q}^{\dot\alpha a}_{-\frac{1}{2}}\right).
\end{align}
Hence, the supercharges which survive the orbifold action are $Q^{\alpha\dot1}_{-\frac{1}{2}}$ for $\sigma=+1$ or $\overline{Q}^{\dot\alpha\dot2}_{+\frac{1}{2}}$ for $\sigma=-1$.
We will use one of them to compute the SCI in the next section.

\section{4d $\N=2^*$ Instantons from a 2d superconformal index computation}\label{Sec:2star}
\label{sec:2Star_Instantons}

In this section we warm up for our main calculation that we perform in the next section by reproducing the well known instanton partition function of $\mathcal{N}=2^*$ via a 2d superconformal index (SCI) calculation.
We parameterise our partition function and use a supercharge that survives the orbifold projection \eqref{Qorbifold1} \eqref{Qorbifold2} so that we are well prepared for the next section.

As discussed in the introduction, since the class $\mathcal{S}_k$ gauge theories of interest may be realised within Type II string theory as a theory living on the worldvolume of D$p$ branes with coordinates $X^1,\dots,X^{p+1}$, one of the most important tools we plan to use in this paper is the relationship between the ADHM construction of instantons \cite{Atiyah:1978ri} and D$(p-4)$ branes \cite{Douglas:1996uz,Douglas:1995bn,Witten:1994tz,Tong:2005un,Polchinski:1996na} in other words
\begin{equation}
\text{$|K|$ (A)SD instantons in a D$p$-brane}\equiv\text{$|K|$ (anti-)D$(p-4)$-branes}\,.
\end{equation}
(Anti-)Self-dual ((A)SD) instantons are solutions to the (A)SD Yang-Mills equations $F=\pm\star F$.
 The instanton number 
$
K=\frac{1}{4\pi^2}\int_{\mathcal{M}_4}\tr F\wedge F\in\Z
$
is a topological invariant.
For SD instantons $F=+\star F$ $K\geq0$ while for ASD instantons $F=-\star F$ $K\leq0$. 
Since parity maps $K\to-K$ we can choose to focus only on ASD instantons, corresponding to $\sigma=\sign K=-1$. The moduli space of ASD instantons for the gauge theory living on the D$p$ branes, $\mathcal{M}_{K}^{\text{D$p$}}$, is then isomorphic to the Higgs branch of the theory living on the D$(p-4)$ branes
\begin{equation}
\mathcal{M}_{K}^{\text{D$p$}}\iso \mathcal{M}_{\text{Higgs}}^{\text{$K$ D$(p-4)$}}=\left\{X^{p+2}=X^{p+3}=\dots=X^{10}=0,\mathcal{V}_{p-3}=0\right\}/U(K)
\end{equation}
where $\mathcal{V}_{p-3}=F\bar{F}+ {1\over 2}D^2$ is the scalar potential of the $(p-3)$d next to maximal supersymmetric gauge theory living on the worldvolume of the D$(p-4)$ branes. The vanishing of $F$- and $D$- terms translate into the ADHM constraints \cite{Dorey:2000zq,Dorey:2002ik}.
When supersymmetry is present the Higgs branch is protected from quantum corrections and the fluctuation determinants in the instanton measure cancel. The action of the theory on the D$(p-4)$ branes is the equivalent to the instanton action, hence the partition function of the theory of $K$ D$(p-4)$ branes is then nothing else but the partition function of $K$ instantons (up to a possible overall factor $\Zc_{\text{extra}}$) for the gauge theory living on the D$p$ branes
\begin{align}
\Zc_{\text{extra}}\Zc^{\text{D$p$}}_{K\text{-inst}}(a,m,\dots)&=\int_{\mathcal{M}_{K,r}^{\text{D$p$}}}e^{-S_{\text{inst}}(a,m,\dots,\mu)}d\mu
=\Zc^{\text{$K$ D$(p-4)$}}_{\text{Higgs}}(a,m,\dots)\,. 
\end{align}
The factor $\Zc_{\text{extra}}$ is often present due to the fact that the theory on the D$(p-4)$ branes provides the UV completion of the ADHM sigma model \cite{Hwang:2014uwa,Kim:2012gu} and therefore it may contain extra degrees of freedom which do not appear in the ADHM construction. Those extra degrees of freedom generally decouple from the the ADHM degrees of freedom and the partition function factorises as above.
The case that interests us is the case $p=5$, i.e. D$5$ branes on $\mathbb{R}^4\times T^2$, thus we have to compute the partition function of the 2d gauge theory living on the world volume D$1$ branes wrapping a $T^2$. This partition function is the 2d superconformal index a.k.a. flavoured elliptic genus.

\subsection{D1 worldvolume theory}
Before discussing the supersymmetric index we must first discuss the worldvolume theory living on the D$1$ branes in the low energy limit in the presence of the D5s.
\paragraph{D1-D1}
The theory arising from quantising open strings stretching between $K$ parallel and coincident D$p$-branes is given by $p+1$ dimensional Yang-Mills theory with $16$ supercharges, for $p=1$ that is the well known $\N=(8,8)$ SYM theory. In terms of multiplets under the $\N=(4,4)$ subalgebra given by $\overline{Q}_{+\frac{1}{2}}^{\dot\alpha \dot a},\overline{Q}_{-\frac{1}{2}}^{\dot\alpha a}$ they form a $\N=(4,4)$ vector multiplet $V$ and hypermultiplet $H$, which can be thought of as the reduction to 2d of a 4d $\N=2$ vector multiplet and hypermultiplet respectively. $V$ contains a 2d gauge field $A_{\pm}$, four scalars degrees of freedom $Y_{a\dot a}$, right moving fermions $\overline{\lambda}^{\dot\alpha a}_{+\frac{1}{2}}$ and left moving fermions $\overline{\xi}_{-\frac{1}{2}}^{\dot\alpha\dot a}$. $H$ contains scalars $X_{\alpha\dot\alpha}$, right moving fermions $\xi_{+\frac{1}{2}}^{\alpha\dot a}$ and left moving fermions $\lambda^{\alpha a}_{-\frac{1}{2}}$.

\paragraph{D1-D5}
Open D$1$-D$5$ strings preserve $\N=(4,4)$ supersymmetry and gives rise to a $\N=(4,4)$ hypermultiplet $U$ in the bifundamental representation of $U(K)\times SU(N)$. $U$ contains two complex scalars $\phi^{\dot\alpha}$ and their conjugates $\phi^{\dagger}_{\dot\alpha}$, and fermions $\chi^{\dot a}_{+\frac{1}{2}}$, $\psi^{a}_{-\frac{1}{2}}$ plus their conjugates $\chi^{\dagger}_{+\frac{1}{2}\dot{a}}$, $\psi^{\dagger}_{-\frac{1}{2}a}$ .
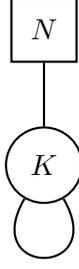
\begin{figure}
\centering
\begin{tikzpicture}[square/.style={regular polygon,regular polygon sides=4},thick,scale=0.9]
 \node(G) at (0,0)[circle,draw,minimum size=1cm]{$K$};
 \node(F) at (0,2)[square,draw]{$N$};
 \draw (F.-90)--(G.90);
 \draw (G) to [out=240,in=300,looseness=6](G);
\end{tikzpicture}
\caption{\it The $\N=(4,4)$ 2d quiver of the gauge theory on $K$ D1 branes in the presence of $N$ D5s. Using $\N=(4,4)$ notation, solid lines denote hypermultiplets, while the circular node denotes the $U(K)$ vector multiplet.}
\label{fig:D1quiver}
\end{figure}
Finally, the field content may be conveniently summarised in the quiver diagram of Figure \ref{fig:D1quiver} .

\subsection{The 2d index calculation}

We now turn to the computation of the supersymmetric index a.k.a flavoured elliptic genus partition function for our $\N=(4,4)$ theory.
The supersymmetric index can be understood as the Witten index of the theory quantised on $\mathbb{S}^1\times \mathbb{R}$ refined by fugacities which keep track of further relevant quantum numbers and it is independent of the coupling constants of the theory.
 Since our theory admits a free field limit, computing the SCI is a `letter counting' problem on $\mathbb{R}^2$ in the radial quantisation  \cite{Putrov:2015jpa,Gadde:2013wq,Gadde:2013ftv,Gadde:2014ppa,Gadde:2013lxa,Nakayama:2011pa,Cordova:2017ohl}. 
For theories with a Lagrangian description the index  can also be obtained using localisation techniques \cite{Benini:2013nda,Benini:2013xpa} and explicitly performing the path integral of the 2d theory on $T^2$, however for simplicity we will follow the former approach.

We also choose to view our $\N=(4,4)$ theory as an $\N=(0,2)$ theory with additional flavour symmetry. We choose the $\N=(0,2)$ supercharges to be 
\begin{equation}\label{eqn:02susycharges}
Q:=\overline{Q}_{+\frac{1}{2}}^{\dot2\dot1}, \quad\widetilde{Q}:=\overline{Q}_{+\frac{1}{2}}^{\dot1\dot2}\,.
\end{equation} 
2d $\N=(0,2)$ theories have a single right moving $U(1)_{\mathfrak{R}}$ $R$-symmetry. The $\N=(0,2)$ IR $R$-symmetry for this model was computed in \cite{Tong:2014yna} and it is given by\footnote{Our D$5$/D$1$ setup is precisely that of \cite{Tong:2014yna} with $Q_5^+=N$, $Q_5^-=0$, $R^-=-2J_R$ and $R^+=-2J_R^R$}
\begin{equation}
\label{eq:IRrsymmetry}
\mathfrak{R}_{\text{IR}}=-2J_R\,,
\end{equation}
under which $\mathfrak{R}_{\text{IR}}[Q]=-1$ and $\mathfrak{R}_{\text{IR}}[\widetilde{Q}]=+1$ 

We will compute the index which counts cohomology classes of $\widetilde{Q}$. Since $\widetilde{Q}$ and its conjugate\footnote{We wish to stress that also $\widetilde{Q}^{\dagger}=\widetilde{S}$ is preserved by the orbifold that we will impose in the next section. To see this note that  $\widetilde{S}$ is `uncharged' under the orbifold generators \eqref{Qorbifold1} and \eqref{Qorbifold2}.} $\widetilde{Q}^{\dagger}=\widetilde{S}$ commute with $SU(2)_{\alpha},SU(2)_{a}$ we may include fugacities $v,w$ for their Cartans. Furthermore, they also commute with the diagonal subgroup $SU(2)_D\subset SU(2)_{\dot\alpha}\times SU(2)_{\dot a}$ hence we also include a fugacity $z$ for its Cartan $J_D=J_R+J^R_R$.
Recall that\footnote{The Cartans of $\mathfrak{su}(2)_{a}$ and $\mathfrak{su}(2)_{\dot a}$ denoted as $J^R_{L}$ and $J_R^R$ can be written in terms of the generators $J_{710}$ and $J_{89}$ which are the $U(1)$ rotations in the respective planes  \eqref{eq:RsymmetryRotations}.} the Cartans of $\mathfrak{su}(2)_{\alpha},\mathfrak{su}(2)_{\dot\alpha},\mathfrak{su}(2)_{a}$ and $\mathfrak{su}(2)_{\dot a}$  all commute with the orbifold and the fugacities $v,w$ and $z$ that we introduced here will still be meaningful for our calculation in the next section.
 We also include fugacities $x_A$ for the Cartans $f_A$ of $\mathfrak{su}(N)$ and $y_I$ for the Cartans $g_I$ of $\mathfrak{u}(K)$. The Witten index is then defined as  
\begin{equation}
\label{eqn:2starindex}
\Zc^{6d,N}_K(q,t,s,p,x_A)= \Tr\left[(-1)^Fq^{H_-}v^{2J_L}w^{2J_L^R}z^{2J_R+2J_R^R}\prod_{A=1}^Nx_A^{f_A}\prod_{I=1}^Ky_{I}^{g_I}\right]
\end{equation}
where $F=F_-+F_+$ is the fermion number, $H_-$, $H_+$ are the left and right moving UV Hamiltonians respectively generating $\mathfrak{iso}(T^2)$. In Euclidean signature we define $2H_{\pm}=H\mp\iu P$ and $q:=e^{2\pi\iu\tau}$ with $\tau$ the complex structure of the $T^2$ is generated by $w\sim w+1\sim w+\tau$. Explicitly, we will work with the square torus with complex structure $\tau=\iu\beta_6/\beta_5$ with $\beta_5$, $\beta_6$ the radii of the two $\mathbb{S}^1$ factors.
In radial quantisation the conformal map from the plane to the cylinder is $Z_{56}=e^{2\pi\iu w}$, where $w:=\sigma+\iu t$ and Lorentz transformations $Z_{56}\mapsto e^{\iu\theta}Z_{56}$ are then mapped to translations around the $\mathbb{S}^1$ factor of the cylinder
\begin{equation}\label{lorentz}
\left(\sigma,t\right)\mapsto \left(\sigma+\frac{\theta}{2\pi}, t\right)
\end{equation}
generated by $P=i J_{56}$. 

One of the crucial properties of the Witten index \eqref{eqn:2starindex} is that it receives contributions only from those states which satisfy 
\begin{equation}
\label{eq:delta}
\delta= \left\{ \widetilde{Q} , \widetilde{S} \right\} =H_+-\frac{1}{2}\mathfrak{R}_{\text{UV}}
=0
\end{equation}
where $\mathfrak{R}_{\text{UV}}$ is the $\N=(0,2)$ $R$-symmetry at the UV fixed point and the index is hence independent of $\overline{q}$. Furthermore the index \eqref{eqn:2starindex} is also independent of all continuous parameters such as coupling constants and Fayet-Iliopoulos parameters \cite{Witten:1982df,Kinney:2005ej}, hence we can compute the index in the free field limit where it reduces to a counting problem.

In the free field limit we have a $\N=(0,2)$ superconformal theory with  $\mathfrak{Vir}\oplus\overline{\mathfrak{sVir}}_{\N=2,NS}$ symmetry where $\mathfrak{Vir}$ is the standard $(\N=0)$ left-moving Virasoro algebra generated by $\left\{L_n,c\right\}$, $\overline{\mathfrak{sVir}}_{\N=2,NS}$ is the $\N=2$ super-Virasoro algebra in the NS sector generated by $\left\{\overline{L}_n,\overline{G}^{\pm}_r,\overline{J}_n,\overline{c}\right\}$ and $n,r+\frac{1}{2}\in\mathbb{Z}$. Our choice of the Neveu-Schwarz basis over the Ramond basis is purely for calculational convenience and the index is independent of this choice up to an overall factor \cite{Cordova:2017ohl}. We will require the following brackets of the $\overline{\mathfrak{sVir}}_{\N=2,NS}$ algebra:
\begin{equation}
\begin{gathered}
\left\{\overline{G}^+_r,\overline{G}^-_s\right\}=\overline{L}_{r+s}+\frac{1}{2}(r-s)\overline{J}_{r+s}+\frac{\overline{c}}{6}\left(r^2-\frac{1}{4}\right)\delta_{r+s,0}\,,\\
\left[\overline{L}_0,\overline{G}^{\pm}_{r}\right]=-r\overline{G}^{\pm}_{r}\,,\quad \left[\overline{J}_0,\overline{G}^{\pm}_{r}\right]=\pm\overline{G}^{\pm}_{r}\,.
\end{gathered}
\end{equation}
In the free field limit, where it is appropriate to refer to the index as the SCI, we identify 
\begin{equation}
\begin{gathered}
L_0=H_-,\quad \overline{L}_0=H_+,\quad \overline{J}_0=\mathfrak{R}_{\text{UV}} 
\\
Q=\overline{G}^{-}_{-\frac{1}{2}},\quad S=\overline{G}^{+}_{+\frac{1}{2}},\quad \widetilde{Q}=\overline{G}^{+}_{-\frac{1}{2}},\quad \widetilde{S}=\overline{G}^{-}_{+\frac{1}{2}} \, .
\end{gathered}
\end{equation}
Away from the free limit, the theory is not conformal and we have an RG flow from the free UV fixed point to an IR fixed point. The $R$-charge assignments generally change along RG flow. 
Nonetheless, the index is RG invariant and we can evaluate the index at the IR fixed point by using the non-anomalous $R$-symmetry assignment in the IR which, in our case, is \eqref{eq:IRrsymmetry}.

At the UV fixed point the shortening condition  \eqref{eq:delta} can be written as
\begin{equation}\label{eqn:bpscondition}
\delta=\left\{\overline{G}^+_{-\frac{1}{2}},\overline{G}^-_{+\frac{1}{2}}\right\}=\overline{L}_{0}-\frac{1}{2}\overline{J}_{0}=0\,.
\end{equation}
and the states contributing to the index must have $\overline{J}_0=2\overline{L}_0$, so we can trivially write $L_0=\frac{1}{2}\overline{J}_0+L_0-\overline{L}_0$. To account for the change of the $R$-symmetry assignment along the RG flow we have that $\overline{J}_0=\mathfrak{R}_{\text{UV}}\to \mathfrak{R}_{\text{IR}}$ and hence going to the IR is taken into account by shifting 
\begin{equation}
q^{L_0}=  q^{L_0-\overline{L}_0+\frac{1}{2}\mathfrak{R}_\text{UV}}  \to q^{L_0-\overline{L}_0+\frac{1}{2}\mathfrak{R}_\text{IR}}   \,.
\end{equation}
while holding $L_0$, $\overline{L}_0$ fixed to their values in the UV.
In words, we evaluate the trace over the local operators at the UV fixed point, however, since the index is a renormalisation group invariant, we evaluate the SCI of the IR fixed point by using the non-anomalous $R$-symmetry in the IR \eqref{eq:IRrsymmetry}.  

\subsubsection{Letter counting}
As stressed earlier the index may be computed in the free field limit. This is done by identifying all `letters' with $\delta=0$. The single letter partition functions for the $\N=(4,4)$ multiplets may be easily read off from Tables \ref{tab:Vletters}, \ref{tab:Hletters}, \ref{tab:Uletters} in Appendix \ref{App:EllipticGenus}. They are given by
\begin{align}
&i_V(q,w,z,y_I)=\left[\frac{\left(w+w^{-1}\right)\left(z+qz^{-1}\right)-qz^{-2}-z^2-2q}{1-q}\right]\sum_{I,J=1}^Ky_Iy_J^{-1}\,,\label{eqn:letterfV}\\
&i_H(q,v,w,z,y_I)=\left[\frac{q^{\frac{1}{2}}\left(v+v^{-1}\right)\left(z+z^{-1}-w^{-1}-w\right)}{1-q}\right]\sum_{I,J=1}^Ky_Iy_J^{-1}\,,\label{eqn:letterfH}\\
&i_U(q,w,z,x_A,y_I)=\left[\frac{q^{\frac{1}{2}}\left(z+z^{-1}-w^{-1}-w\right)}{1-q}\right]\sum_{I=1}^K\sum_{A=1}^N\left(y_Ix_A^{-1}+y_I^{-1}x_A\right)\label{eqn:letterfU}\,.
\end{align}
The full index is then by enumerating all possible `words' and then projecting onto gauge invariant states by integrating over the Haar measure $[dG]$ of the $G=U(K)$ gauge group. $[dU\left(K\right)]$ may be reduced to an integral over the maximal torus $T(G)$ at the cost of introducing the Vandermonde determinant $\prod_{I\neq J}\left(y_I-y_J\right)$
\begin{equation}
\oint\left[dU\left(K\right)\right]=\frac{1}{K!}\oint_{T(G)}\prod_{I=1}^K\frac{dy_I}{2\pi\iu}\prod_{I\neq J}\left( y_I-y_J\right)\,.
\end{equation}
The contour is taken over $|y_I|=1$ . The full index is then given by
\begin{equation}
\Zc^{6d,N}_K(q,v,w,z,x_A)=\oint\left[dU\left(K\right)\right]\Zc^{(0)}(q,v,w,z,x_A,y_I)\prod_{\mathcal{M}=V,H,U}\mathcal{Z}_{\mathcal{M}}(q,v,w,z,x_A,y_I)
\end{equation}
where $\Zc^{(0)}$ is the Casimir contribution which may, apriori, depend on all fugacities. It is given by \cite{Benini:2011nc,Kim:2009wb,Imamura:2011su,Rastelli:2016tbz,Bobev:2015kza,Assel:2015nca,Ardehali:2015bla,Kim:2012ava}
\begin{equation}\label{eqn:casimir}
\Zc^{(0)}(q,v,w,z,x_A,y_I)=q^{\frac{1}{2}E_{\text{Casimir}}},\quad E_{\text{Casimir}}=\substack{\text{Finite}\\q\to1}\left[\sum_{\mathcal{M}}\frac{\partial i_{\mathcal{M}}}{\partial\log q}\right]
\end{equation}
and 
\begin{equation}\label{eqn:singpartfunction}
\mathcal{Z}_{\mathcal{M}}(q,v,w,z,x_A,y_I):=\PE\left[i_{\mathcal{M}}(q,v,w,z,x_A,y_I)\right],\quad \PE\left[i(\cdot)\right]:=\exp\left[\sum_{n=1}^{\infty}\frac{1}{n}i(\cdot^n)\right]
\end{equation}
where $i_{\mathcal{M}}$ are the single letter partition functions \eqref{eqn:letterfV}, \eqref{eqn:letterfH} and \eqref{eqn:letterfU}. Explicitly:
\begin{align}
&\Zc_V(q,w,z,y_I)=\prod_{I,J=1}^K\frac{\left(q\frac{y_I}{y_J};q\right)^2\theta\left(qz^{-2}\frac{y_I}{y_J};q\right)}{\theta\left(wz\frac{y_I}{y_J};q\right)\theta\left(qwz^{-1}\frac{y_I}{y_J};q\right)}\,,\\
&\Zc_H(q,v,w,z,y_I)=\prod_{I,J=1}^K\frac{\theta\left(q^{\frac{1}{2}}vw\frac{y_I}{y_J};q\right)\theta\left(q^{\frac{1}{2}}v^{-1}w\frac{y_I}{y_J};q\right)}{\theta\left(q^{\frac{1}{2}}vz^{-1}\frac{y_I}{y_J};q\right)\theta\left(q^{\frac{1}{2}}v^{-1}z^{-1}\frac{y_I}{y_J};q\right)}\,,\\
&\Zc_U(q,w,z,x_A,y_I)=\prod_{I=1}^K\prod_{A=1}^N\frac{\theta\left(q^{\frac{1}{2}}w\frac{x_A}{y_I};q\right)\theta\left(q^{\frac{1}{2}}w\frac{y_I}{x_A};q\right)}{\theta\left(q^{\frac{1}{2}}z^{-1}\frac{x_A}{y_I};q\right)\theta\left(q^{\frac{1}{2}}z^{-1}\frac{y_I}{x_A};q\right)}\,,
\end{align}
where
\begin{equation}
\theta\left(x;q\right):=\left(x;q\right)\left(qx^{-1};q\right),\quad \left(x;q\right):=\prod_{n=0}^{\infty}\left(1-xq^n\right)=\exp\left[-\sum_{n=1}^{\infty}\frac{1}{n}\frac{x^n}{1-q^n}\right] 
\end{equation}
is the $q$-theta function and $q$-Pochhammer symbol respectively. They are related to the Jacobi theta function
\begin{equation}\label{eqn:Jacobitheta}
\theta_1(x;q):=\iu q^{\frac{1}{12}}\eta(q)(x^{-\frac{1}{2}}-x^{\frac{1}{2}})\prod_{n=1}^{\infty}(1-xq^n)(1-x^{-1}q^n)=\iu q^{\frac{1}{12}}\eta(q)x^{-\frac{1}{2}}\theta(x;q)\,,
\end{equation}
where $\eta(q):=q^{\frac{1}{24}}\prod_{n=1}^{\infty}(1-q^n)=q^{\frac{1}{24}}(q;q)$ is the Dedekind eta function. Finally, we conclude that the full index is given by
\begin{equation}\label{eqn:6dpartitionfunctionK}
\begin{aligned}
&\Zc^{6d,N}_K=\frac{\left(q;q\right)^{2K}}{K!}\oint_{T(G)}\prod_{I=1}^K\frac{dy_I}{2\pi\iu y_I}\Zc^{(0)}(q,\dots)\prod_{I=1}^K\prod_{A=1}^N\frac{\theta\left(q^{\frac{1}{2}}w\frac{x_A}{y_I};q\right)\theta\left(q^{\frac{1}{2}}w\frac{y_I}{x_A};q\right)}{\theta\left(q^{\frac{1}{2}}z^{-1}\frac{x_A}{y_I};q\right)\theta\left(q^{\frac{1}{2}}z^{-1}\frac{y_I}{x_A};q\right)}\\
&\times\prod_{I\neq J}\theta\left(\frac{y_I}{y_J};q\right)\prod_{I,J=1}^K\frac{\theta\left(qz^{-2}\frac{y_I}{y_J};q\right)\theta\left(q^{\frac{1}{2}}vw\frac{y_I}{y_J};q\right)\theta\left(q^{\frac{1}{2}}v^{-1}w\frac{y_I}{y_J};q\right)}{\theta\left(wz\frac{y_I}{y_J};q\right)\theta\left(qwz^{-1}\frac{y_I}{y_J};q\right)\theta\left(q^{\frac{1}{2}}vz^{-1}\frac{y_I}{y_J};q\right)\theta\left(q^{\frac{1}{2}}v^{-1}z^{-1}\frac{y_I}{y_J};q\right)}
\end{aligned}
\end{equation}
where we used the identity $\left(x;q\right)=\left(1-x\right)\left(qx;q\right)$. It also useful to assemble the quantity
\begin{equation}\label{eqn:6dpartitionfunction}
\Zc^{6d,N}(q,v,w,z,x_A;\mathbf{q}_{6d}):=\sum_{K\geq0}\mathbf{q}^K_{6d}\Zc^{6d,N}_K(q,v,w,z,x_A)
\end{equation}
with $\mathbf{q}_{6d}$ a formal dimensionless parameter. When considering our 6d theory on $\mathbb{R}^4\times \mathbb{S}_5^1\times \mathbb{S}_6^1$ as a 5d theory on $\mathbb{R}^4\times \mathbb{S}_5^1$ dressed by KK modes along  $\mathbb{S}_6^1$ we may regard $\mathbf{q}_{6d}$ as a fugacity for the topological $U(1)$ global symmetry associated to the conserved current $\star_{5d}J=\frac{1}{8\pi^2}\tr F\wedge F$.

\subsection{The 6d instanton partition function}
The countour integrals \eqref{eqn:6dpartitionfunctionK} may be computed via the Jefferey-Kirwan residue prescription \cite{1993alg.geom..7001J,Benini:2013nda,Benini:2013xpa}. The function $\theta\left(y;q\right)$ has simple zeros for $y=q^{a+b/\tau}$ for $a,b\in\Z$ and no poles. Furthermore, to compute residues note that
\begin{equation}
\frac{\partial}{\partial y}\theta\left(y;q\right)|_{y=1}=-\left(q;q\right)^2\,.
\end{equation}
Using the identity $\theta\left(qx;q\right)=\frac{-1}{x}\theta\left(x;q\right)$ the residue is given by
\begin{equation}
\oint_{y=q^{a+b/\tau}}\frac{dy}{2\pi\iu y}\frac{1}{\theta\left(y;q\right)}=(-1)^{a+1}\left(q;q\right)^{-2}q^{\frac{a}{2}\left(a-1\right)}\,.
\end{equation}
Hence, we can perform the residue prescription. The integrand of \eqref{eqn:6dpartitionfunction} has simple poles at
\begin{gather}
y_I=y_J\left(zw\right)^{\pm1},\quad y_I=y_J\left(\frac{z}{qw}\right)^{\pm1},\\
y_I=y_J\left(\frac{vz}{q^{1/2}}\right)^{\pm1},\quad y_I=y_J\left(\frac{z}{vq^{1/2}}\right)^{\pm1},\quad y_I=x_A\left(\frac{z}{q^{1/2}}\right)^{\pm1}.\label{eqn:polekeep}
\end{gather}
As explained in \cite{Kim:2011mv,Kim:2012gu} only residues arising from the poles \eqref{eqn:polekeep} should be kept. We assume that the $x_A$'s are sufficiently generic and furthermore we close the contour such that we collect residues coming from poles with the positive sign exponents. 
The solutions to \eqref{eqn:polekeep} may be classified by $N$-coloured Young's diagrams $\vec{Y}=\left\{Y_1,\dots,Y_N\right\}$ with each diagram $Y_A$ containing $|Y_A|$ boxes such that $|\vec{Y}|:=\sum_{A}|Y_A|=K$. Given a Young's diagram $Y_A$ a box $s$ is labelled by coordinates $(l,p)$ and the corresponding pole is given by
\begin{equation}
y(s)=x_A\left(\frac{z}{q^{1/2}}\right)^{l+p-1}v^{l-p} \, .
\end{equation}
The residue for a fixed coloured Young diagram is then
\begin{equation}
\Zc^{6d,N}_{\vec{Y}}=\Zc^{(0)}\prod_{A,B=1}^N\prod_{s\in Y_A}\frac{\theta\left(q^{-1}zw^{-1}\mathcal{E}_{AB};q\right)\theta\left(zw\mathcal{E}_{AB};q\right)}{\theta\left(\mathcal{E}_{AB};q\right)\theta\left(q^{-1}z^2\mathcal{E}_{AB};q\right)}
\, ,
\end{equation}
where we defined
\begin{equation}
\mathcal{E}_{BA}:=\frac{x_B}{x_A}\left(\frac{vz}{q^{1/2}}\right)^{L_A(s)}\left(\frac{q^{1/2}v}{z}\right)^{A_B(s)+1}
\end{equation}
and where $L_B(s)$ and $A_B(s)$ denote the distance from the box $s$ to the right end and the bottom of the Young diagram $Y_B$ respectively. $\Zc^{(0)}=q^{\frac{1}{2}E_{\text{Casimir}}}$ is the Casimir contribution \eqref{eqn:casimir}. To compute it one is forced to specify the $q$-dependence of the fugacities, we hence define  
\begin{equation}\label{eqn:5dvariables}
v:=q^{\frac{\beta_5\epsilon_-}{2\iu\pi}}\,,\quad wq^{\frac{1}{2}}:=q^{\frac{\beta_5m}{\iu\pi}}\,,\quad zq^{-\frac{1}{2}}:=q^{\frac{\beta_5\epsilon_+}{2\iu\pi}}\,,\quad x_A:=q^{\frac{\beta_5a_A}{\iu\pi}}\,,
\end{equation}
where we used the shorthand notation
\begin{equation}
\epsilon_{\pm}:=\epsilon_1\pm\epsilon_2\,.
\end{equation}
$\Zc^{(0)}$ is then a constant and is given by
\begin{equation}
\Zc^{(0)}=q^{\frac{\beta_5^2NK}{\pi^2}\left(\frac{\epsilon_+}{2}-m+\frac{\iu\pi}{\beta_5}\right)\left(\frac{\epsilon_+}{2}+m\right)}\,.
\end{equation}
The $K$ instanton partition function \eqref{eqn:6dpartitionfunctionK} is then given by summing over all coloured Young diagrams $\vec{Y}$.
Hence, equation \eqref{eqn:6dpartitionfunction} finally reads
\begin{equation}\label{eqn:6dpartitionfunctionyng}
\Zc^{6d,N}:=\sum_{K\geq0}\mathbf{q}_{6d}^K\sum_{\vec{Y}\atop{|\vec{Y}|=K}}\Zc_{\vec{Y}}^{6d,N}=\sum_{\vec{Y}}\mathbf{q}^{|\vec{Y}|}_{6d}\Zc_{\vec{Y}}^{6d,N}\,.
\end{equation}

\subsection{The 5d limit of the instanton partition function}
Reducing the D5/D1 system on $\mathbb{S}^1_6$ by taking $\beta_6\to0$ results in 5d $\N=2$ SYM with gauge group $SU(N)$ at Chern-Simons level $\kappa=0$. Hence by either taking the 5d limit directly to \eqref{eqn:6dpartitionfunctionyng} or taking the limit directly the contour integral \eqref{eqn:6dpartitionfunctionK} we expect to obtain the instanton partition function for the 5d theory. Here we take the first approach but we detail the limit of the contour integral expression in Appendix \ref{5dLimitOfUnorbifolded}.

Recall that $q=e^{2\pi\iu\tau}$ and $\tau=\iu\beta_6/\beta_5$ therefore this limit corresponds to taking
\begin{equation}
q\to1\,.
\end{equation}
To take the limit, first note that the ratio of $q$-theta function may be rewritten as
\begin{equation}
\frac{\theta\left(q^a;q\right)}{\theta\left(q^b;q\right)}=\frac{\left[a\right]_q}{\left[b\right]_q}\prod_{n=1}^{\infty}\frac{\left[n+a\right]_q\left[n-a\right]_q}{\left[n+b\right]_q\left[n-b\right]_q}
\end{equation}
where $[n]_q:=(1-q^n)/(1-q)$ is the $q$-number. The $q$-number has the property that \begin{equation}
\lim_{q\to1}[n]_q=n
\end{equation}
and therefore
\begin{equation}\label{eqn:thetafunctionlimit}
\lim_{q\to1}\frac{\theta\left(q^a;q\right)}{\theta\left(q^b;q\right)}=\frac{\sinh\iu\pi a}{\sinh\iu\pi b}\,.
\end{equation}
Further note that by definition 
\begin{equation}
\lim_{q\to1}\Zc^{(0)}=q^{\frac{1}{2}E_{\text{Casimir}}}=1\,.
\end{equation}
Applying this to \eqref{eqn:6dpartitionfunctionyng} yields
\begin{align}
\Zc^{5d,N}=&\sum_{K\geq0}\mathbf{q}_{5d}^K\sum_{\vec{Y}\atop{|\vec{Y}|=K}}\prod_{A,B=1}^N\prod_{s\in Y_A}\frac{\sinh\beta_5\left(E_{AB}+\frac{\epsilon_+}{2}-m\right)\sinh\beta_5\left(E_{AB}+\frac{\epsilon_+}{2}+m\right)}{\sinh\beta_5\left(E_{AB}\right)\sinh\beta_5\left(E_{AB}+\epsilon_+\right)}\label{5dpartitionfunctionyng}\\
=&\sum_{\vec{Y}}\mathbf{q}_{5d}^{|\vec{Y}|}\Zc_{\vec{Y}}^{5d,N}
\end{align}
where we defined
\begin{equation}
E_{BA}=a_B-a_A+\epsilon_1L_A(s)-\epsilon_2\left(A_B(s)+1\right)\,,
\end{equation}
such that $\mathcal{E}_{AB}=q^{\frac{E_{AB}}{\iu\pi}}$ and we take $\mathbf{q}_{5d}=\lim_{q\to1}\mathbf{q}_{6d}$.
 This reproduces the instanton partition for the mass deformed 5d $\N=2$ theory (a.k.a. $\N=1^*$) on $\mathbb{R}^4\times \mathbb{S}^1_5$ which was computed via localisation of the path integral of the ADHM quantum mechanics in e.g. \cite{Hwang:2014uwa,Kim:2011mv,Hori:2014tda}.
Hence we indeed identify
\begin{equation*}
\Zc^{5d,N}=\Zc^{5d,N}_{\text{inst},\N=1^*}.
\end{equation*}

\subsection{The 4d limit of the instanton partition function}
Armed with the above, the equivalence of $\Zc^{5d,N}$ in the 4d limit $(\beta_5\to0)$ with the instanton partition function $\Zc^{4d,N}_{\text{inst},\N=2^*}$ for the 4d $\N=2^*$ theory is essentially trivial to prove. Taking the $\beta_5\to0$ limit of \eqref{5dpartitionfunctionyng} or equivalently evaluating the contour integrals \eqref{eqn:4dpartitionfunction} we obtain
\begin{align}
\Zc^{4d,N}=&\sum_{K\geq0}\mathbf{q}_{4d}^K\sum_{\vec{Y}\atop{|\vec{Y}|=K}}\prod_{A,B=1}^N\prod_{s\in Y_A}\frac{\left(E_{AB}+\frac{\epsilon_+}{2}-m\right)\left(E_{AB}+\frac{\epsilon_+}{2}+m\right)}{E_{AB}\left(E_{AB}+\epsilon_+\right)}\\
=&\sum_{\vec{Y}}\mathbf{q}_{4d}^{|\vec{Y}|}\Zc_{\vec{Y}}^{4d,N}=\Zc^{4d,N}_{\text{inst},\N=2^*}
\end{align}
we then identify $m$ as the hypermultiplet mass in the $\Omega$-background \cite{Okuda:2010ke} and set $\mathbf{q}_{4d}=\lim_{\beta_5\to0}\mathbf{q}_{5d}$.
It may be shown \cite{Bruzzo:2002xf,Shadchin:2005mx,Shadchin:2005cc} that the vector multiplet contribution is given by
\begin{equation}\label{eqn:zvec}
z_{\text{vec}}(a,\vec{Y})=\prod_{A,B=1}^N\prod_{s\in Y_A}\frac{1}{E_{AB}\left(E_{AB}+\epsilon_+\right)}
\end{equation}
and the contribution from the adjoint hypermultiplet is
\begin{equation}
z_{\text{adj}}(a,m,\vec{Y})=\prod_{A,B=1}^N\prod_{s\in Y_A}\left(E_{AB}+\frac{\epsilon_+}{2}-m\right)\left(E_{AB}+\frac{\epsilon_+}{2}+m\right)
\end{equation}
also note that
\begin{equation}
z_{\text{adj}}\left(a,\frac{\epsilon_+}{2},\vec{Y}\right)=\frac{1}{z_{\text{vec}}(a,\vec{Y})}.
\end{equation}

\section{Orbifolding to 4d $\N=2$/$\N=1$  circular/toroidal quivers}
\label{sec:Orbifolding}

The main goal of this paper is to compute the 2d index in the presence of the $\Gamma=\mathbb{Z}_{\ell}\times\mathbb{Z}_k$ orbifold before reducing to the zero dimensional matrix model partition function which is expected to be equal to the partition function of instantons for the $\Gamma$-orbifolded 4d theory.

In principle we could work directly with the 2d orbifolded theory by working out the projections and writing down the Lagrangian and computing it's partition function. However we prefer instead to work with the SCI interpreted as a counting device to which we implement projection onto $\Gamma$-invariant states. 

We take the same approach, as one takes when computing, e.g. the supersymmetric Lens space $L(1,r)\times \mathbb{S}^1$ indices \cite{Benini:2011nc,Razamat:2013jxa,Razamat:2013opa,Alday:2013rs,Hikida:2006qb}. This is also the same method one uses to compute orbifold partition functions in 2d CFT \cite{francesco1996conformal}.

\subsection{Orbifolding the supersymmetric index}\label{subsec:orbindex}
If we denote some (mother) theory by $M$ we may obtain a new (daughter) theory $D=M/\Gamma$ by quotienting out by an orbifold group $\Gamma$ which is generically embedded inside both the global symmetry group $F_M$ and gauge group $G_M$ of $M$. We collectively denote the generators of $\Gamma$ by $\gamma$. If $M$ is a supersymmetric theory with a supercharge $\mathcal{Q}$, then it is possible to count cohomology classes of $\mathcal{Q}$, i.e. to compute the supersymmetric index of $M$ for the supercharge $\mathcal{Q}$ which, providing that $M$ admits a suitable free field limit such that standard letter counting techniques can be applied, is schematically defined to be
\begin{equation}
\mathcal{I}_M(a)=\Tr_{\mathcal{H}_M}\left[(-1)^Fe^{-\beta\left\{\mathcal{Q},\mathcal{Q}^{\dagger}\right\}}a^{f_M}\right]
\end{equation}
where $f_M$ collectively denotes the subset of linearly independent generators of $F_M$ such that $\left[\mathcal{Q},f_M\right]=0$ and $a$ their fugacities.
If we assume that $\Gamma$ is abelian and furthermore commutes with both $\mathcal{Q}$ and its conjugate 
\begin{equation}
[\mathcal{Q},\gamma]=[\mathcal{Q}^{\dagger},\gamma]=0
\end{equation}
then the theory $D$ will also generically possesses at least one supersymmetry, namely $\mathcal{Q}$. $D$ also has reduced global and gauge symmetry groups $F_D=C_{F_M}\left(\Gamma\right)$, $G_D=C_{G_M}\left(\Gamma\right)$ where $C_{G}(S)$ denotes the centraliser of $S$ in $G$ which, of course, depends on the choice of embedding $\Gamma\hookrightarrow F_M\times G_M$.
One may then obtain the supersymmetric index of $D$ for the supercharge $\mathcal{Q}$ by means of projection:
\begin{equation}\label{eqn:ID}
\mathcal{I}_D(a)=\sum_{\rho}\Tr_{\mathcal{H}_{\rho}}\left[\int[d\Gamma]\varepsilon^{\gamma}(-1)^Fe^{-\beta\left\{\mathcal{Q},\mathcal{Q}^{\dagger}\right\}}a^f\right]\,.
\end{equation}
The orbifold projection can be considered as a gauging of the $\Gamma$ symmetry of the theory $M$ resulting in a new theory $D$. 
The `integral' over the invariant Haar measure of the group $\Gamma$ implements the projection onto $\Gamma$-invariant states. When $\Gamma$ is discrete and abelian the Haar measure is simply given by summing over all elements of the group and dividing by the number of elements of the group
\begin{equation}
\int[d\Gamma]=\frac{1}{|\Gamma|}\sum_{\varepsilon\in\Gamma}\,.
\end{equation}
Since $\mathcal{H}_M$ is a Hilbert space with grading by global symmetries, it may be decomposed $\mathcal{H}_M=\oplus_{\rho}\mathcal{H}_{\rho}$ according to the $\Gamma$ action. To include states which may also be twisted in the `time' direction we must also sum over different vacuua $\mathcal{H}_{\rho}$. This definition automatically receives contributions from both untwisted sectors as well as sectors which may be twisted by global or gauge symmetries. Note that in the computation of $\mathcal{I}_D$, since $G_M$ was gauged, one should `integrate' over all independent (up to $G/C_{G_M}\left(\Gamma\right)$ gauge transformations) embeddings $\Gamma\hookrightarrow G_M$. Note also that in some cases one may also choose to instead use a weighted sum over embeddings, for example discrete theta angles \cite{Razamat:2013opa}. On the other hand $F_M$ is not gauged and hence one should fix a particular embedding $\Gamma\hookrightarrow F_M$ which in turn specifies the global symmetry of the daughter theory $D$.
\subsubsection{A toy example - the orbifold index of a free Fermi multiplet}
Let us proceed with a simple toy example. Staying in 2d, since that is the most relevant for us, we let $M$ be the theory of a free $\N=(0,2)$ Fermi multiplet $\psi_{-\frac{1}{2}},\overline{\psi}_{-\frac{1}{2}}$. Both left moving fermions have $\overline{L}_0=\overline{J}_0=0$ and hence both satisfy the BPS condition $\delta=\left\{\mathcal{Q},\mathcal{Q}^{\dagger}\right\}=0$ \eqref{eqn:bpscondition} furthermore they both have $L_0=\frac{1}{2}$. The free Fermi multiplet admits a $U(1)_{f}$ flavour symmetry generated by $f$ under which $\psi_{-\frac{1}{2}}$, $\overline{\psi}_{-\frac{1}{2}}$ have charges $f=+1,-1$. The total global symmetry is then $F_M=U(1)_f \times \mathrm{Iso}(T^2)\times U(1)_{\mathfrak{R}}$. In particular $\mathbb{R}^2\subset\mathfrak{iso}(T^2)$ is the algebra of translations around the two cycles of the torus, where $\overline{L}_0-L_0$ generates radial translations and $\overline{L}_0+L_0$ generate time translations. $U(1)_{\mathfrak{R}}$ is the $R$-symmetry generated by $\overline{J}_0=\mathfrak{R}$ under which $\mathcal{Q},\mathcal{Q}^{\dagger}$ have charges $\mathfrak{R}=-1,+1$ as before. 
Enumerating the letters of Table \ref{tab:freefermi} we obtain 
\begin{equation}
\mathcal{I}_M(a,q)=\Tr_{\mathcal{H}_M}\left[(-1)^Fq^{L_0}a^{f}\right]=\PE\left[-\frac{q^{\frac{1}{2}}\left(a+a^{-1}\right)}{1-q}\right]=\theta\left(aq^{\frac{1}{2}};q\right)
\end{equation}
\begin{table}[htb]
\centering
 \begin{tabular}{|c|c|c|c|c|c|c|} 
 \hline
 Letter & $L_0$& $\overline{L}_0$& $\overline{J}_0$ & $f$ & Index&Orb Index\\\hline
\hline
$\psi_{-\frac{1}{2}}$&$1/2$&$0$ & $0$ & $+1$& $q^{\frac{1}{2}}a$&$-q^{1/2}a$\\\hline
$\overline\psi_{-\frac{1}{2}}$&$1/2$ &$0$ & $0$ & $-1$& $q^{1/2}a^{-1}$&$-\varepsilon^{-1}q^{1/2}a^{-1}$\\\hline
\hline
$\partial_{-}$ &$1$ &$0$& $0$ &$0$& $q$&$\varepsilon^{-1}q$\\\hline
\end{tabular}
\caption{\it Letters of the Fermi multiplet.}
\label{tab:freefermi}
\end{table}
We consider the theory obtained by the $\mathbb{Z}_k$ orbifold $D=M/\mathbb{Z}_k$ where $\mathbb{Z}_k$ has an action inside all three factors of $F_M$. To preserve the supercharge $\mathcal{Q}$ we require that $\mathbb{Z}_k$ translation in $\mathrm{Iso}(T^2)$ and rotation by $U(1)_{\mathfrak{R}}$ acts by equal but opposite amounts. Hence the $\mathbb{Z}_k$ group elements are of the form 
\begin{equation}
 e^{\frac{2\pi\iu}{k}\gamma}\in\mathbb{Z}_k\,,\quad\gamma=\frac{f}{2}+\left(\overline{L}_0-L_0\right)+\frac{\mathfrak{R}}{2}\,.
\end{equation}
Since the quotient acts in the radial direction only $\mathcal{I}_D$ is simply obtained by taking the Pleythistic exponent of the orbifolded single letter index
\begin{align}
i(a,q)=&\frac{1}{k}\sum_{\varepsilon\in\mathbb{Z}_k}\left[-\left(\varepsilon^{-\frac{1}{2}}q^{\frac{1}{2}}\varepsilon^{\frac{1}{2}}a+\varepsilon^{-\frac{1}{2}}q^{\frac{1}{2}}\varepsilon^{-\frac{1}{2}}a^{-1}\right)\right]\sum_{n\geq0}\varepsilon^{-n}q^n\\
=&\frac{1}{k}\sum_{\varepsilon\in\mathbb{Z}_k}\left[-q^{\frac{1}{2}}a-q^{-\frac{1}{2}}a^{-1}\right]\sum_{\tilde{n}\geq0}q^{k\tilde{n}}\sum_{j=1}^{k-1}\varepsilon^{-j}q^j+q^{-\frac{1}{2}}a^{-1}\\
=&-\frac{aq^{\frac{1}{2}}+a^{-1}q^{k-\frac{1}{2}}}{1-q^k}
\end{align}
where we used the basic fact that $\sum\limits_{\varepsilon\in\mathbb{Z}_k}\varepsilon=0$. The index for the theory $D$ for the supercharge $\mathcal{Q}$ is then
\begin{equation}
\mathcal{I}_D(a,q)=\PE\left[\mathit{i}(a,q)\right]=\PE\left[-\frac{aq^{\frac{1}{2}}+a^{-1}q^{k-\frac{1}{2}}}{1-q^k}\right]=\theta\left(aq^{\frac{1}{2}};q^k\right)\,.
\end{equation}
We now move to apply this general discussion to the D$1$ worldvolume theory.
\subsection{Computing the orbifolded superconformal index}
The orbifold acts on the coordinates $X^1,\dots,X^{10}$ as in \eqref{eqn:orbifoldactionR10}. The orbifold action is also embedded in the $SU(\ell kN)$ flavour group as in \eqref{eqn:orbifoldactionSUN}, where we also scaled $N\to \ell kN$ with respect to the previous section. Furthermore, the orbifold also has an action inside the $U(K)$ gauge group of the D$1$ worldvolume theory breaking $U(K)\to\prod_{n=1}^{\ell}\prod_{i=1}^kU(K_{ni})$, $\sum_{n,i}K_{ni}=K$ and $U(0)$ is defined to be the trivial group. As in equation \eqref{eqn:orbifoldactionSUN} the action may be conjugated to an element of the maximal torus $g\in T(U(K))$
\begin{equation}\label{eqn:orbifoldactionUK}
g=\diag\left(\omega_{\ell}\omega_k\mathbb{I}_{K_{11}},\dots,\omega_{\ell}\omega_k^k\mathbb{I}_{K_{1k}},\dots,\omega^{\ell}_{\ell}\omega_k\mathbb{I}_{K_{\ell 1}},\dots, \omega_{\ell}^{\ell}\omega^k_k\mathbb{I}_{K_{\ell k}}\right)\,.
\end{equation}
The only difference is that we do not fix $K_{ni}$ but rather `integrate' over all possible $K_{ni}$ satisfying $\sum_{n,i}K_{ni}=K$ which indeed are in one-to-one correspondence with embeddings $\Gamma\hookrightarrow U(K)$ up to gauge transformations. In the language of the discussion in Section \ref{subsec:orbindex} we consider $G_M=U(K)$, $F_M=SU\left(\ell kN\right)\times \mathrm{Iso}\left(T^2\right)\times Spin(4)_R$ and $\Gamma=\mathbb{Z}_{\ell}\times\mathbb{Z}_k$. Then $C_{SU(\ell kN)}\left(\Gamma\right)$\footnote{Note that here $C_{SU(\ell kN)}\left(\Gamma\right)$ is equivalent to the Levi subgroup $\mathbb{L}$ specified by $\ell$ copies of \eqref{eqn:levipartition}.} coincides with $SU(N)^{\ell k}$. On the other hand $C_{G_M}(\Gamma)=\prod_{n,i}U(K_{ni})$ corresponding to the (unordered) partition $K=K_{11}+\dots+K_{\ell k}$ but since $G_M$ was gauged we sum over all partitions.

For convenience we choose to split the Cartans $f_A$, $A=1,\dots,\ell kN$ of $\mathfrak{su}(\ell kN)$ into Cartans $f_{ni,A}$, $A=1,\dots,N$ of $\oplus_{ni}\mathfrak{su}(N)$ we also do the same with the $\mathfrak{u}(K)$ Cartan generators $g_I$, $I=1,\dots, K$ into Cartans $g_{ni,I}$, $I=1,\dots,K_{ni}$ of $\oplus_{ni}\mathfrak{u}(K_{ni})$. Following the above discussion, recalling that the supercharge $\widetilde{Q}$ (given in \eqref{eqn:02susycharges}) and its conjugate $\widetilde{S}$ commute with the orbifold action, we compute
\begin{equation}\label{eqn:orbindex}
\begin{aligned}
\Zc^{6d,\ell,k}_K(q,v,w,z,x_{ni,A})=\Tr&\left[\frac{1}{\ell k}\sum_{\varepsilon_{\ell}\in\Z_{\ell}\atop{\varepsilon_{k}\in\Z_{k}}}\varepsilon_{\ell}^{\gamma_{\ell}}\varepsilon_k^{\gamma_k}\prod_{n,i}\prod_{A=1}^N\left(\varepsilon_{\ell}^n\varepsilon_k^i\right)^{f_{ni,A}}\prod_{I=1}^{K_{ni}}\left(\varepsilon_{\ell}^n\varepsilon_k^i\right)^{g_{ni,I}}\right.\\
&\phantom{[}\times\left.\vphantom{\sum_{\varepsilon_{\ell}\in\Z_{\ell}\atop{\varepsilon_{k}\in\Z_{k}}}}(-1)^Fq^{H_-}v^{2J_L}w^{2J_L^R}z^{2J_R+2J_R^R}\prod_{n,i}\prod_{A=1}^Nx_{ni,A}^{f_{ni,A}}\prod_{I=1}^{K_{ni}}y_{ni,I}^{g_{ni,I}}\right]\,,
\end{aligned}
\end{equation}
where the first line corresponds to the projection operator $\int\left[d\Gamma\right]\varepsilon^{\gamma}$ of \eqref{eqn:ID} implementing the Douglas-Moore orbifold procedure with: 
\begin{equation}
\gamma_{\ell}=J_{710}-J_{89}:=2J_L^R \,,\quad \gamma_k:=J_{56}-J_{89}=J_{56}+J_L^R-J_R^R\,.
\end{equation}
Recall that in computing the index previously we mapped the plane to the cylinder $Z_{56}=e^{2\pi\iu\left(\sigma+\iu t\right)}$ hence, rotations of the plane $Z_{56}\mapsto e^{\iu\theta}Z_{56}$ are mapped to translations $\sigma\mapsto\sigma+\frac{\theta}{2\pi}$ \eqref{lorentz}. Hence, quotienting out by rotations on the plane corresponds, after the conformal map, to quotienting out translations generated by $\overline{L}_0-L_0$ on the torus.

We assume that the IR $R$-symmetry $\mathfrak{R}_{\text{IR}}$ does not change under the orbifold; in which case, relegating the explicit derivation to Appendix \ref{App:EllipticGenus}, we obtain the orbifolded single letter indices which are denoted by $i_{\Gamma V}$, $i_{\Gamma H}$, $i_{\Gamma U}$ and are given by equations \eqref{eqn:orbletterfV}, \eqref{eqn:orbletterfH}, \eqref{eqn:orbletterfU} respectively. In addition, for a fixed partition $\{K_{ni}\}$ the Haar measure becomes
\begin{equation}
\oint\left[dU(K)\right]\to\prod_{n=1}^{\ell}\prod_{i=1}^k\frac{1}{K_{ni}!}\oint_{T\left[U(K_{ni})\right]}\prod_{I=1}^{K_{ni}}\frac{dy_{ni,I}}{2\pi\iu y_{ni,I}}\prod_{I\neq J}\left(1-\frac{y_{ni,I}}{ y_{ni,J}}\right):=\oint\prod_{n,i}\left[dU(K_{ni})\right]
\end{equation}
which coincides with the Haar measure for the product group $\prod_{n,i}U(K_{ni})$. The `orbifolded' index for a fixed partition $\{K_{ni}\}$ is then
\begin{equation}\label{eqn:indexfixedmn}
\Zc^{6d,N,\ell,k}_{\{K_{ni}\}}(q,v,w,z,x_{ni,A}):=\oint\prod_{n,i}\left[dU\left(K_{ni}\right)\right]\Zc^{(0)}_{\{K_{ni}\}}(q,\dots)\prod_{\mathcal{M}= V,H,U}\mathcal{Z}_{\Gamma\mathcal{M}}(q,\dots)\,.
\end{equation}
 The Casimir contribution $\Zc^{(0)}_{\{K_{ni}\}}$ is defined in the same way as \eqref{eqn:casimir}. The `orbifolded' single letter partition function are defined as in \eqref{eqn:singpartfunction} and are explicitly given by
\begin{equation}
\begin{aligned}\Zc_{\Gamma V}&=\left(q^k;q^k\right)^{2K}\prod_{n,i}\prod_{I\neq J}\frac{\theta\left(\frac{y_{ni,I}}{y_{ni,J}};q^k\right)}{\left(1-\frac{y_{ni,I}}{y_{ni,J}}\right)}\prod_{i\neq j}\prod_{I=1}^{K_{ni}}\prod_{J=1}^{K_{nj}}\theta\left(q^{L_{ij}}\frac{y_{ni,I}}{y_{nj,J}};q^k\right)\\
&\times\prod_{n,i,j}\frac{\prod_{I=1}^{K_{ni}}\prod_{J=1}^{K_{nj}}\theta\left(z^{-2}q^{L_{ij}+1}\frac{y_{ni,I}}{y_{nj,J}};q^k\right)}{\prod_{I=1}^{K_{ni}}\prod_{J=1}^{K_{(n+1)j}}\theta\left(wzq^{L_{ij}}\frac{y_{ni,I}}{y_{(n+1)j,J}};q^k\right)\theta\left(z^{-1}wq^{L_{ij}+1}\frac{y_{ni,I}}{y_{(n+1)j,J}};q^k\right)}\,,
\end{aligned}
\end{equation}
\begin{equation}
\Zc_{\Gamma H}=\prod_{n,i,j}\frac{\prod_{I=1}^{K_{ni}}\prod_{J=1}^{K_{(n+1)j}}\theta\left(wvq^{L_{ij}+\frac{1}{2}}\frac{y_{ni,I}}{y_{(n+1)j,J}};q^k\right)\theta\left(wv^{-1}q^{L_{ij}+\frac{1}{2}}\frac{y_{ni,I}}{y_{(n+1)j,J}};q^k\right)}{\prod_{I=1}^{K_{ni}}\prod_{J=1}^{K_{nj}}\theta\left(vz^{-1}q^{L_{ij}+\frac{1}{2}}\frac{y_{ni,I}}{y_{nj,J}};q^k\right)\theta\left(v^{-1}z^{-1}q^{L_{ij}+\frac{1}{2}}\frac{y_{ni,I}}{y_{nj,J}};q^k\right)}\,,
\end{equation}
\begin{equation}
\Zc_{\Gamma U}=\prod_{n,i,j}\prod_{A=1}^N\prod_{I=1}^{K_{ni}}\frac{\theta\left(wq^{L_{ij}+\frac{1}{2}}\frac{y_{ni,I}}{x_{(n+1)j,A}};q^k\right)\theta\left(w^{-1}q^{k-L_{ji}-\frac{1}{2}}\frac{y_{ni,I}}{x_{(n-1)j,A}};q^k\right)}{\theta\left(z^{-1}q^{L_{ij}+\frac{1}{2}}\frac{y_{ni,I}}{x_{nj,A}};q^k\right)\theta\left(zq^{k-L_{ji}-\frac{1}{2}}\frac{y_{ni,I}}{x_{nj,A}};q^k\right)}\,.
\end{equation}
Here 
\begin{equation}\label{eqn:Ljv}
L_{ij}:=\left\{
\#\in\mathbb{Z} \,\middle|\text{ $0\leq \#\leq k-1$ and $\#=i-j\bmod k$}\right\}
\end{equation} 
is a unique integer and is equivalent to the $L_{ij}=[[i-j]]$ as defined in \cite{Benini:2011nc}. It satisfies the important relation 
\begin{equation}\label{eqn:usefulid}
L_{ij}=\begin{cases}
k-L_{ji} & \text{$i-j\neq0\bmod k$}\\
0&\text{$i-j=0\bmod k$}
\end{cases}
\end{equation} 
It also satisfies $L_{(i+k)j}=L_{ij}=L_{i(j+k)}$ allowing us to consistently abuse the orbifold condition $i\sim i+k$ within products, etc.

We also consider a rewriting of \eqref{eqn:indexfixedmn} in which many simplifications become manifest. We change integration variables and define shifted variables
\begin{equation}\label{eqn:gaugetrans}
y_{ni,I}\to q^{k-i}y_{n,\mathcal{I}}\,,\quad x_{ni,A}:=q^{k-i}\tilde{x}_{n,\mathcal{A}}\,,
\end{equation} 
we also combined the indices such that $\mathcal{A}=iA=1,\dots,kN$ and $\mathcal{I}=iI=1,\dots,K_n$ where $K_n:=\sum_{i=1}^kK_{ni}$.

The shifts may be interpreted in the following way: since the effect of a non-trivial holonomy may always be locally removed by a gauge transformation, the shift of the fugacities $x$ can be thought of as a gauge transformation on the D$5$ theory. On the other hand, the shift of the $y$'s may always be made by a change of integration variables.

Those variables allow us to make rewritings of the form
\begin{align}
\prod_{i,j=1}^k\theta\left(q^{L_{ij}-i+j}x;q^k\right)&=\theta\left(x;q^k\right)^k\prod_{i>j}\theta\left(q^{L_{ij}-i+j}x;q^k\right)\prod_{j>i}\theta\left(q^{L_{ij}-i+j}x;q^k\right)\\
&=\theta\left(x;q^k\right)^k\prod_{i>j}\theta\left(x;q^k\right)\prod_{j>i}\theta\left(q^{k}x;q^k\right)\\
&=\prod_{i,j=1}^k\theta\left(x;q^k\right)\prod_{j>i}\left(\frac{-1}{x}\right)
\end{align}
where the second line follows by applying the definition \eqref{eqn:Ljv} and the relation \eqref{eqn:usefulid} while the third line is courtesy of the identity $\theta\left(qx;q\right)=\frac{-1}{x}\theta\left(x;q\right)$. In terms of those variables we have
\begin{equation}
\begin{aligned}\label{eqn:lkmnpartfnrewrite}
\Zc^{6d,N,\ell,k}_{\{K_{ni}\}}&(q,v,w,z,x_{n,iA})=\left(q^k;q^k\right)^{2K}\prod_{n=1}^{\ell}\left(\prod_{i=1}^k\frac{1}{K_{ni}!}\right)\prod_{n=1}^{\ell}\prod_{j>i}\prod_{A=1}^N\left(\frac{\tilde{x}_{n+1,jA}\tilde{x}_{n-1,jA}}{\tilde{x}_{n,jA}^2}\right)^{K_{ni}}\\
&\times\prod_{n=1}^{\ell}\oint\prod_{I=1}^{K_{n}}\frac{dy_{n,\mathcal{I}}}{2\pi\iu y_{n,\mathcal{I}}}\Zc^{(0)}_{\{K_{ni}\}}\prod_{\mathcal{A}=1}^{kN}\prod_{\mathcal{I}=1}^{K_{n}}\frac{\theta\left(wq^{\frac{1}{2}}\frac{y_{n,\mathcal{I}}}{\tilde{x}_{n+1,\mathcal{A}}};q^k\right)\theta\left(w^{-1}q^{-\frac{1}{2}}\frac{y_{n,\mathcal{I}}}{\tilde{x}_{n-1,\mathcal{A}}};q^k\right)}{\theta\left(z^{-1}q^{\frac{1}{2}}\frac{y_{n,\mathcal{I}}}{\tilde{x}_{n,\mathcal{A}}};q^k\right)\theta\left(zq^{-\frac{1}{2}}\frac{y_{n,\mathcal{I}}}{\tilde{x}_{n,\mathcal{A}}};q^k\right)}\\
&\times\prod_{n=1}^{\ell}\prod_{\mathcal{I}\neq \mathcal{J}}\theta\left(\frac{y_{n,\mathcal{I}}}{y_{n,\mathcal{J}}};q^k\right)\prod_{\mathcal{I},\mathcal{J}=1}^{K_{n}}\frac{\theta\left(z^{-2}q\frac{y_{n,\mathcal{I}}}{y_{n,\mathcal{J}}};q^k\right)}{\theta\left(vz^{-1}q^{\frac{1}{2}}\frac{y_{n,\mathcal{I}}}{y_{n,\mathcal{J}}};q^k\right)\theta\left(v^{-1}z^{-1}q^{\frac{1}{2}}\frac{y_{n,\mathcal{I}}}{y_{n,\mathcal{J}}};q^k\right)}\\
&\times\prod_{\mathcal{I}=1}^{K_n}\prod_{\mathcal{J}=1}^{K_{n+1}}\frac{\theta\left(wvq^{\frac{1}{2}}\frac{y_{n,\mathcal{I}}}{y_{n+1,\mathcal{J}}};q^k\right)\theta\left(wv^{-1}q^{\frac{1}{2}}\frac{y_{n,\mathcal{I}}}{y_{n+1,\mathcal{J}}};q^k\right)}{\theta\left(wz\frac{y_{n,\mathcal{I}}}{y_{n+1,\mathcal{J}}};q^k\right)\theta\left(z^{-1}wq\frac{y_{n,\mathcal{I}}}{y_{n+1,\mathcal{J}}};q^k\right)}\,,
\end{aligned}
\end{equation}

The full `orbifolded' index \eqref{eqn:orbindex} is then given by summing over all partitions $\{K_{ni}\}$ of $K$. However, in analogy with \eqref{eqn:6dpartitionfunction} from the 5d point of view we expect to have a $U(1)^{\ell k}$ topological symmetry associated to the currents $\star_{5d}J_{ni}=\frac{1}{8\pi^2}\tr F_{ni}\wedge F_{ni}$ for the $(n,i)$\textsuperscript{th} gauge node in the quiver. Since the D-instantons serve as sources for those currents and the associated instanton number $K_{ni}$ is related to the partition $\{K_{ni}\}$ following our discussion in Section \ref{subsec:orbindex} we may weight each contribution with fugacities $\mathbf{q}_{6d,ni}$ for each current and assemble the quantity
\begin{equation}
\label{eqn:lkpartitionfunction}
\Zc^{6d,N,\ell,k}(q,v,w,z,x_{n,\mathcal{A}};\mathbf{q}_{6d,ni})=\sum_{K\geq0}\sum_{\{K_{ni}\}\atop\sum_{n,i}K_{ni}=K}\left(\prod_{n,i}\mathbf{q}_{6d,ni}^{K_{ni}}\right)\Zc^{6d,\ell,k}_{\{K_{ni}\}}(q,v,w,z,x_{ni,A})
\end{equation}
the sum over all possible partitions of $K$ is equivalent to summing over all embeddings $\Gamma\hookrightarrow U(K)$ up to gauge transformation.

\subsection{The 6d orbifolded instanton partition function}
In this section we check that for $k=1$ our partition function indeed reproduces the known result for the partition function for $\{K_1,\dots,K_{\ell}\}$ D1-branes in the presence of $\ell N$ D5 branes on $A_{\ell-1}$ singularity as computed in \cite{Gadde:2015tra}. Furthermore we show that the $\mathbb{Z}_k$ orbifolded index may be obtained from the partition function without orbifold (up to an overall shift in the Casimir contribution) with the substitution rule $x\to\tilde{x}$ and $q\to q^k$.

We must first evaluate the contour integrals \eqref{eqn:indexfixedmn}. It bares many resemblances with the unorbifolded $(\ell=k=1)$ partition function \eqref{eqn:6dpartitionfunction} discussed in Section \ref{Sec:2star}. In particular, the poles are now located at
\begin{gather}
y_{n+1,\mathcal{I}}=y_{n,\mathcal{J}}\left(zw\right),\quad y_{n-1,\mathcal{I}}=y_{n,\mathcal{J}}\left(\frac{z}{qw}\right),\\
y_{n,\mathcal{I}}=y_{n,\mathcal{J}}\left(\frac{vz}{q^{1/2}}\right)^{\pm1},\quad y_{n,\mathcal{I}}=y_{n,\mathcal{J}}\left(\frac{z}{vq^{1/2}}\right)^{\pm1},\quad y_{n,\mathcal{I}}=\tilde{x}_{n,\mathcal{A}}\left(\frac{z}{q^{1/2}}\right)^{\pm1}.\label{eqn:polekeeporb}
\end{gather}

Proceeding in an analogous way to the unorbifolded index we again assume that the $\tilde{x}$'s can be made sufficiently generic and that the correct residues to collect are again those coming from the poles \eqref{eqn:polekeeporb}. Solutions to \eqref{eqn:polekeeporb} are then in fact classified by $\ell$ lots of $kN$-coloured Young's diagrams which we label by $\vec{Y}_n=\{Y_{n,\mathcal{A}}\}=\left\{Y_{n,1},\dots,Y_{n,kN}\right\}$ again with each diagram $Y_{n,\mathcal{A}}$ containing $|Y_{n,\mathcal{A}}|$ boxes such that $|\vec{Y}_n|:=\sum_{\mathcal{A}}|Y_{n,\mathcal{A}}|=K_n$ where $K_n:=\sum_{i=1}^kK_{ni}$ as before. Given a Young's diagram $Y_{n,\mathcal{A}}$ a box $s$ is labelled by coordinates $(l,p)$ and the corresponding pole is given by
\begin{equation}
y_n(s)=\tilde{x}_{n,\mathcal{A}}\left(\frac{z}{q^{1/2}}\right)^{l+p-1}v^{l-p}
\end{equation}
Hence, for a fixed partition $\{K_{ni}\}$ the residue of \eqref{eqn:indexfixedmn} for a fixed set of $kN$-tuples $\vec{Y}_1,\dots,\vec{Y}_{\ell}$ is
\begin{equation}\label{eqn:Nlkresidue}
\begin{aligned}
\Zc^{6d,N,\ell,k}_{\{K_{ni}\},\{\vec{Y}_n\}}&=\Zc^{(0)}_{\{K_{ni}\},\{\vec{Y}_n\}}\prod_{n=1}^{\ell}\left(\frac{K_n!}{\prod_{j=1}^kK_{ni}!}\right)\prod_{n=1}^{\ell}\prod_{j>i}\prod_{A=1}^N\left(\frac{\tilde{x}_{n+1,jA}\tilde{x}_{n-1,jA}}{\tilde{x}_{n,jA}^2}\right)^{K_{ni}}\\
&\times\prod_{n=1}^{\ell}\prod_{\mathcal{A},\mathcal{B}=1}^{kN}\frac{\prod_{s\in Y_{n+1,\mathcal{A}}}\theta\left(q^{-1}zw^{-1}\mathcal{E}_{(n+1)n,\mathcal{A}\mathcal{B}};q^k\right)\prod_{s\in Y_{n,\mathcal{A}}}\theta\left(zw\mathcal{E}_{n(n+1),\mathcal{A}\mathcal{B}};q^k\right)}{\prod_{s\in Y_{n,\mathcal{A}}}\theta\left(\mathcal{E}_{nn,\mathcal{A}\mathcal{B}};q^k\right)\prod_{s\in Y_{n,\mathcal{A}}}\theta\left(q^{-1}z^2\mathcal{E}_{nn,\mathcal{A}\mathcal{B}};q^k\right)}
\end{aligned}
\end{equation}
where we defined
\begin{equation}\label{eqn:Eip}
\mathcal{E}_{nm,AB}:=\frac{\tilde{x}_{n,\mathcal{A}}}{\tilde{x}_{m,\mathcal{B}}}\left(\frac{vz}{q^{1/2}}\right)^{L_{m,\mathcal{B}}(s)}\left(\frac{q^{1/2}v}{z}\right)^{\left(A_{n,\mathcal{A}}(s)+1\right)}
\end{equation}
where $L_{n,\mathcal{A}}(s)$ and $A_{n,\mathcal{A}}(s)$ denote the distance from the box $s$ to the right end and the bottom of the Young diagram $Y_{n,\mathcal{A}}$ respectively.
Furthermore, we also notice the multinomial coefficient $K_n!/\prod_{i=1}^kK_{ni}!=\binom{\sum_i K_{ni}}{K_{n1},\dots,K_{nk}}$.

We are still yet to describe the Casimir contribution. Again we must specify the $q$-dependence of the fugacities. We write
\begin{equation}\label{eqn:lk5dvariables}
v:=q^{\frac{\beta_5k\epsilon_-}{2\iu\pi}},\quad wq^{\frac{1}{2}}:=q^{\frac{\beta_5km}{\iu\pi}},\quad zq^{-\frac{1}{2}}:=q^{\frac{\beta_5k\epsilon_+}{2\iu\pi}},\quad \tilde{x}_{n,\mathcal{A}}:=q^{\frac{\beta_5k\tilde{a}_{n,\mathcal{A}}}{\iu\pi}}\,.
\end{equation}

The Casimir contribution is explicitly given in equation \eqref{eqn:lkCasimir} and after evaluating its residue for a fixed $kN$-tuples $\{\vec{Y}_n\}$ is
\begin{equation}
\begin{aligned}
&\Zc^{(0)}_{\{K_{ni}\},\{\vec{Y}_n\}}=\prod_{n=1}^{\ell}\prod_{\mathcal{A}}^{kN}q^{\frac{kK_n\beta_5^2}{2\pi^2}\left[2\tilde{a}_{n,\mathcal{A}}^2-\tilde{a}_{n-1,\mathcal{A}}^2-\tilde{a}_{n+1,\mathcal{A}}^2+2m\left(\tilde{a}_{n+1,\mathcal{A}}-\tilde{a}_{n-1,\mathcal{A}}\right)\right]}\\
&\times q^{\frac{-k^2KN\beta_5^2}{\iu\pi}\left(\frac{\epsilon_+}{2}+m\right)\left(\frac{\epsilon_+}{2}-m+\frac{1}{\beta_5\iu\pi}\right)}\prod_{n=1}^{\ell}\prod_{\mathcal{A},\mathcal{B}=1}^{kN}\prod_{s\in Y_{n,\mathcal{B}}}q^{\frac{\beta_5^2}{2\pi^2}\left(\phi_n(s)+\frac{\epsilon_+}{2}\right)\left(\tilde{a}_{n+1,\mathcal{A}}+\tilde{a}_{n-1,\mathcal{A}}-2\tilde{a}_{n,\mathcal{A}}\right)}\,,
\end{aligned}
\end{equation}
where as before we must use the definitions \eqref{eqn:lk5dvariables}. 
For a box $s\in Y_{n,\mathcal{A}}$ the function $\phi_n(s)$ is given by
\begin{equation}
\phi_n(s)=\tilde{a}_{n,\mathcal{A}}+(l-1)\epsilon_1+(p-1)\epsilon_2\,.
\end{equation}
Equation \eqref{eqn:lkmnpartfnrewrite} is then obtained by summing over all $\vec{Y}_n$
\begin{equation}\label{eqn:lkmnpartfnrewriteresi}
\Zc^{6d,N,\ell,k}_{\{K_{ni}\}}(q,v,w,z,x_{ni,A})=\sum_{{\vec{Y_1},\dots,\vec{Y_{\ell}}\atop|\vec{Y}_n|=K_n}}\Zc^{6d,N,\ell,k}_{\{K_{ni}\},\{\vec{Y}_n\}}(q,v,w,z,x_{ni,A})\,.
\end{equation}

\paragraph{The case $k=1$:}
We may immediately compare our expression \eqref{eqn:lkmnpartfnrewriteresi} for $k=1$ to the $\{K_1,\dots,K_{\ell}\}$ partition function computed in \cite{Gadde:2015tra} equation (5.5). After taking the appropriate decoupling limit by `opening up' the quiver we find agreement\footnote{Our parameters are related to those of \cite{Gadde:2015tra} by $t_{\text{them}}=zvq^{1/2}$, $d_{\text{them}}=zv^{-1}q^{-1/2}$, $c_{\text{them}}=q^{1/2}w$.} up to a choice of overall normalisation $\Zc^{(0)}$. 

\paragraph{The case $k>1$:}
By inspection we can immediately see, writing in a schematic fashion and suppressing the unchanged arguments, that
\begin{equation}
\Zc^{6d,N,\ell,k}_{\{K_{ni}\}}\left(q,x\right)=\prod_{n=1}^{\ell}\binom{K_n}{K_{n1},\dots,K_{nk}}\prod_{n=1}^{\ell}\prod_{j>i}\prod_{A=1}^N\left(\frac{\tilde{x}_{n+1,jA}\tilde{x}_{n-1,jA}}{\tilde{x}_{n,jA}^2}\right)^{K_{ni}}\Zc^{6d,kN,\ell,1}_{\{kK_n\}}\left(q^k,\tilde{x}\right)\,.
\end{equation}
In words, we claim that the orbifolded $(k>1)$ index corresponding to the choice or $\{K_{ni}\}$ may be obtained from the unorbifolded $(k=1)$ index by substituting $q\to q^k$, $x_{n,\mathcal{A}}\to\tilde{x}_{n,\mathcal{A}}$ and multiplying by an overall factor.


\subsection{The 5d limit  of the orbifolded instanton partition function}
We can again take the 5d $\beta_6\to0$ $(q\to1)$ limit. For $k=1$ the resulting 5d theories are the $\N=1$ circular quivers denoted by $\mathcal{N}_{N,\ell}$ on $\mathbb{R}^4\times \mathbb{S}^1_5$. For $k>1$ we expect the resulting 5d theory is $\mathcal{N}_{k N,\ell}$ with $k$ codimension $1$ defects which fill the $\mathbb{R}^4$ and are located at points $\Theta=\Theta_{j=1,\dots,k}$ where $\Theta\sim \Theta+2\pi$ is the coordinate of $\mathbb{S}_5^1$.

Following the same procedure as before and making the identifications \eqref{eqn:lk5dvariables}, we find
\begin{equation}\label{eqn:lknm5dpartitionfunctionyng}
\begin{aligned}
\Zc^{5d,N,\ell,k}_{\{K_{ni}\}}=&\sum_{\vec{Y_1},\vec{Y_2},\dots,\vec{Y_{\ell}}\atop{|\vec{Y_n}|=K_n}}\prod_{n=1}^{\ell}\binom{K_n}{K_{n1},\dots,K_{nk}}\\
&\times\prod_{\mathcal{A},\mathcal{B}=1}^{kN}\prod_{s\in Y_{n,\mathcal{A}}}\frac{\sinh\beta_5\left(E_{n(n-1),\mathcal{A}\mathcal{B}}+\frac{\epsilon_+}{2}-m\right)\sinh\beta_5\left(E_{(n-1)n,\mathcal{A}\mathcal{B}}+\frac{\epsilon_+}{2}+m\right)}{\sinh\beta_5\left(E_{nn,\mathcal{A}\mathcal{B}}\right)\sinh\beta_5\left(E_{nn,\mathcal{A}\mathcal{B}}+\epsilon_+\right)}
\end{aligned}
\end{equation}
again $\lim_{q\to1}\Zc^{(0)}_{\{K_{ni}\}}=1$ and the function $E_{nm,\mathcal{A}\mathcal{B}}$ is defined as
\begin{equation}
E_{nm,\mathcal{A}\mathcal{B}}:=\tilde{a}_{n,\mathcal{A}}-\tilde{a}_{m,\mathcal{B}}+\epsilon_1L_{n,\mathcal{A}}(s)-\epsilon_2\left(A_{n,\mathcal{B}}(s)+1\right)\,.
\end{equation}
As before the instantons partition function of the resulting 5d theory is given by
\begin{equation}
\Zc^{5d,N,\ell,k}=\sum_{\{K_{ni}\}}\left(\prod_{n,i}\mathbf{q}_{5d,ni}^{K_{ni}}\right)\Zc^{5d,N,\ell,k}_{\{K_{ni}\}}
\,.
\end{equation}

\subsection{The 4d limit  of the orbifolded instanton partition function}
Finally, we take the 4d $\beta_5\to0$ limit. We expect that by taking the 4d limit we land on the 4d torodial quiver SCFTs in Class $\mathcal{S}_k$. In particular, we want to compare our expression in this limit with the expression proposed in \cite{Mitev:2017jqj}. Applying the 4d limit to \eqref{eqn:lknm5dpartitionfunctionyng} yields:
\begin{equation}\label{eqn:lknm4dpartitionfunctionyng}
\begin{aligned}
\Zc^{4d,N,\ell,k}_{\{K_{ni}\}}=&\sum_{\vec{Y_1},\vec{Y_2},\dots,\vec{Y_{\ell}}\atop{|\vec{Y_n}|=K_n}}\prod_{n=1}^{\ell}\binom{K_n}{K_{n1},\dots,K_{nk}}\\
&\times\prod_{n=1}^{\ell}\prod_{\mathcal{A},\mathcal{B}=1}^{kN}\frac{\prod_{s\in Y_{n+1,\mathcal{A}}}\left(E_{(n+1)n,\mathcal{A}\mathcal{B}}+\frac{\epsilon_+}{2}-m\right)\prod_{s\in Y_{n,\mathcal{B}}}\left(E_{n(n+1),\mathcal{A}\mathcal{B}}+\frac{\epsilon_+}{2}+m\right)}{\prod_{s\in Y_{n,\mathcal{A}}}\left(E_{nn,\mathcal{A}\mathcal{B}}\right)\prod_{s\in Y_{n,\mathcal{A}}}\left(E_{nn,\mathcal{A}\mathcal{B}}+\epsilon_+\right)}
\end{aligned}
\end{equation}
and the partition function of instantons reads
\begin{equation}
\Zc^{4d,N,\ell,k}=\sum_{\{K_{ni}\}}\left(\prod_{n,i}\mathbf{q}_{4d,ni}^{K_{ni}}\right)\Zc^{4d,N,\ell,k}_{\{K_{ni}\}}=\Zc^{4d}_{\text{inst},\tilde{A}_{\ell-1}\times \tilde{A}_{k-1}}\,.
\end{equation}
For $k=1$ \eqref{eqn:lknm4dpartitionfunctionyng} may be compared with the partition function of instantons for the 4d $\N=2$ $\tilde{A}_{\ell-1}$ circular quiver theories.

\subsection{From necklace/toroidal  to linear/cylindrical quivers}

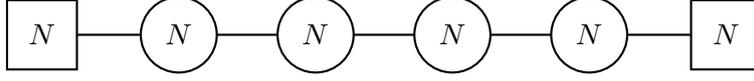
\begin{figure}
\centering
\begin{tikzpicture}[square/.style={regular polygon,regular polygon sides=4},thick,inner sep=0.1em,scale=0.9]
     \draw (0,0) -- (10,0);
     \node[square,draw,minimum size=1.3cm,fill=white] at (0,0) {$N$};
     \node[circle,draw,minimum size=1cm,fill=white] at (2,0) {$N$};
     \node[circle,draw,minimum size=1cm,fill=white] at (4,0) {$N$};
     \node[circle,draw,minimum size=1cm,fill=white] at (6,0) {$N$};
     \node[circle,draw,minimum size=1cm,fill=white] at (8,0) {$N$};
     \node[square,draw,minimum size=1.3cm,fill=white] at (10,0) {$N$};
\end{tikzpicture}
\caption{\it The 5d $\mathcal{N}_{N,\ell}$ quiver for $\ell=5$ after taking the decoupling limit obtained by sending one of the couplings to zero. Similarly, circle reduction in 4d $\mathcal{N}=2$ linear $A_{\ell-1}$ quivers.}
\label{fig:5dNNldecoup}
\end{figure}
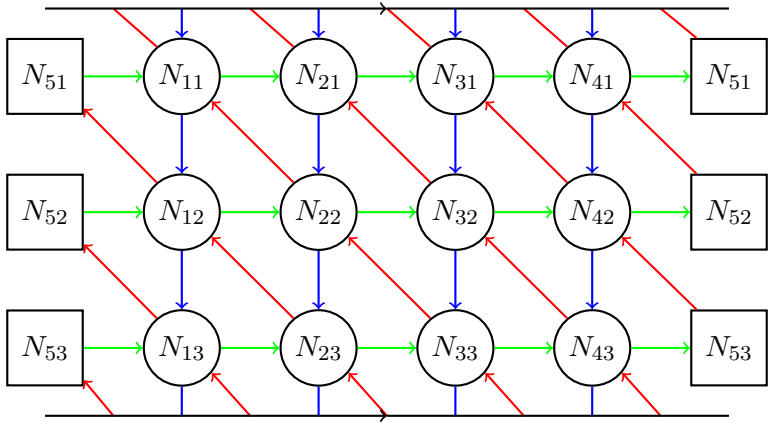
\begin{figure}
\centering
  \begin{tikzpicture}[square/.style={regular polygon,regular polygon sides=4},thick,inner sep=0.1em,scale=0.9]
  
    \node (G11) at (0,0)[circle,draw,minimum size=1cm]{$N_{11}$};
    \node (G21) at (2,0) [circle,draw,minimum size=1cm]{$N_{21}$};
    \node (G31) at (4,0) [circle,draw,minimum size=1cm]{$N_{31}$};
    \node (G41) at (6,0) [circle,draw,minimum size=1cm]{$N_{41}$};
	\node (G51L) at (-2,0)[square,draw,minimum size=1cm]{$N_{51}$};
    \node (G51) at (8,0)[square,draw,minimum size=1cm]{$N_{51}$};
    \node (G12) at (0,-2)[circle,draw,minimum size=1cm]{$N_{12}$};
    \node (G22) at (2,-2) [circle,draw,minimum size=1cm]{$N_{22}$};
    \node (G32) at (4,-2) [circle,draw,minimum size=1cm]{$N_{32}$};
    \node (G42) at (6,-2)[circle,draw,minimum size=1cm]{$N_{42}$};
    \node (G52) at (8,-2)[square,draw,minimum size=1cm]{$N_{52}$};
    \node (G52L) at (-2,-2)[square,draw,minimum size=1cm]{$N_{52}$};
	\node (G13) at (0,-4)[circle,draw,minimum size=1cm]{$N_{13}$};
    \node (G23) at (2,-4) [circle,draw,minimum size=1cm]{$N_{23}$};
    \node (G33) at (4,-4) [circle,draw,minimum size=1cm]{$N_{33}$};
    \node (G43) at (6,-4)[circle,draw,minimum size=1cm]{$N_{43}$};
    \node (G53) at (8,-4)[square,draw,minimum size=1cm]{$N_{53}$};
    \node (G53L) at (-2,-4)[square,draw,minimum size=1cm]{$N_{53}$};
    
    \draw [->,color=blue] (G11.270) to (G12.90);
    \draw [->,color=blue] (G21.270) to (G22.90);
    \draw [->,color=blue] (G31.270) to (G32.90);
    \draw [->,color=blue] (G41.270) to (G42.90);
    \draw [->,color=blue] (G12.270) to (G13.90);
    \draw [->,color=blue] (G22.270) to (G23.90);
    \draw [->,color=blue] (G32.270) to (G33.90);
    \draw [->,color=blue] (G42.270) to (G43.90);
    \draw [->,color=blue] (0,1) to (G11.90);
    \draw [->,color=blue] (2,1) to (G21.90);
    \draw [->,color=blue] (4,1) to (G31.90);
    \draw [->,color=blue] (6,1) to (G41.90);
    \draw [-,color=blue] (G13.270) to (0,-5);
    \draw [-,color=blue] (G23.270) to (2,-5);
    \draw [-,color=blue] (G33.270) to (4,-5);
    \draw [-,color=blue] (G43.270) to (6,-5);
    
    \draw [->,color=green] (G51L.0) to (G11.180);
    \draw [->,color=green](G11.0) to (G21.180);
    \draw [->,color=green] (G21.0) to (G31.180);
    \draw [->,color=green] (G31.0) to (G41.180);
    \draw [->,color=green] (G41.0) to (G51.180);
    \draw [->,color=green] (G52L.0) to (G12.180);
    \draw [->,color=green] (G12.0) to (G22.180);
    \draw [->,color=green] (G22.0) to (G32.180);
    \draw [->,color=green] (G32.0) to (G42.180);
    \draw [->,color=green] (G42.0) to (G52.180);
    \draw [->,color=green] (G53L.0) to (G13.180);
    \draw [->,color=green] (G13.0) to (G23.180);
    \draw [->,color=green] (G23.0) to (G33.180);
    \draw [->,color=green] (G33.0) to (G43.180);
    \draw [->,color=green] (G43.0) to (G53.180);

    \draw [-,color=red] (G11.130) to (-1,1);
    \draw [-,color=red] (G21.130) to (1,1);
    \draw [-,color=red] (G31.130) to (3,1);
    \draw [-,color=red] (G41.130) to (5,1);
    \draw [-,color=red] (G51.130) to (7,1);
    \draw [->,color=red] (G12.130) to (G51L.-40);
    \draw [->,color=red] (G13.130) to (G52L.-40);

	\draw [<-,color=red] (G11.-40) to (G22.130);
    \draw [<-,color=red] (G21.-40) to (G32.130);
    \draw [<-,color=red] (G31.-40) to (G42.130);
    \draw [<-,color=red] (G41.-40) to (G52.130);
    \draw [<-,color=red] (G12.-40) to (G23.130);
    \draw [<-,color=red] (G22.-40) to (G33.130);
    \draw [<-,color=red] (G32.-40) to (G43.130);
    \draw [<-,color=red] (G42.-40) to (G53.130);
    
    \draw [<-,color=red] (G13.-40) to (1,-5);
    \draw [<-,color=red] (G23.-40) to (3,-5);
    \draw [<-,color=red] (G33.-40) to (5,-5);
    \draw [<-,color=red] (G43.-40) to (7,-5);
    \draw [<-,color=red] (G53L.-40) to (-1,-5);

    \draw [->] (-2,1) to (3,1);
    \draw [-] (3,1) to (8,1);
    \draw [->] (-2,-5) to (3,-5);
    \draw [-] (3,-5) to (8,-5);
  \end{tikzpicture}
  \caption{\it The 4d $A_{\ell-1}\times\tilde{A}_{k-1}$ cylindrical quiver in $\N=1$ notation with $\ell=5$, $k=3$ obtained after taking a decoupling limit of the $\tilde{A}_{\ell-1}\times\tilde{A}_{k-1}$ torodial quiver.}
  \label{fig:Skquiverdecoup}
\end{figure}

In this subsection we want to briefly explain, for the sake of the non-expert reader, how we can obtain the instanton partition functions for linear $\N=2$ (generic $\ell$, $k=1$) or cylindrical $\N=1$ quivers (generic $\ell$ and $k$) from the formulas we have just derived that are for necklace $\N=2$ (generic $\ell$, $k=1$) and toroidal $\N=1$ (generic $\ell$ and $k$) quivers respectively.
Firstly, for the  $\N=2$ theories with generic $\ell$ and $k=1$,
we choose a fugacity  $\mathbf{q}_{6d,n}$ for one $n$ corresponding to one coupling constant of the $n^{\text{th}}$ gauge node and send it to zero
in equation
\eqref{eqn:lkpartitionfunction}.
This corresponds to ungauging this gauge factor and breaks the necklace at this node. See Figure \ref{fig:5dNNl} and Figure \ref{fig:5dNNldecoup}. It is useful for notational clarity to ungauge the node with $n=\ell$. Then the hypermultiplets that couple to this node from the left and from the right become fundamental and the Coulomb branch parameters $a_{1}=m_L$ and $a_{\ell}=m_R$ are interpreted as anti-fundamental and fundamental masses, respectively.
Moving on to the toroidal $\N=1$  quivers  with generic $\ell$ and $k$, we can obtain cylindrical  $\N=1$  quivers via ungauging all of the $k$-nodes with  $n=\ell$,
 setting in \eqref{eqn:lkpartitionfunction} $\mathbf{q}_{6d,\ell i}=0$ for all $i=1,\dots,k$. See Figure \ref{fig:Skquiver} and Figure \ref{fig:Skquiverdecoup}. 
 Finally, let us stress that this ungauging procedure can be done for all 6d, 5d and 4d instanton partition functions.

\subsection{From  $\N=2^*$ to $\N=2$ $\tilde{A}_{\ell-1}$ instantons at the orbifold point}
\label{para:2starAl}

In this subsection, we want to  discuss a relationship between instantons of the $\mathcal{N}=2^*$ theory and the $\N=2$ $\tilde{A}_{\ell-1}$ $(k=1)$ at a special point in the parameter space, which is commonly referred to as `the orbifold point'. For the $\tilde{A}_{\ell-1}$ theories this corresponds to setting all instanton parameters equal $\mathbf{q}_{4d,n}\equiv \mathbf{q}$ and the vevs take the special values $a_{n,A}=a_L=a_A$ where $L=nA$, $n=1,\dots,\ell$ and $A=1,\dots ,N$ .

The partition function for the $\tilde{A}_{\ell-1}$ quiver is given by a sum over $\ell$ $N$-coloured Young diagrams with $K_A$ boxes. The idea is that since all the $\mathbf{q}_A$ parameters are equal, we can rewrite the sum 
\begin{equation}
\sum_{\{K_n\}\geq0}\mathbf{q}^{K_1+\dots+K_{\ell}}\sum_{\vec{Y}_{n}\atop|\vec{Y}_{n}|=K_n}=\sum_{K=0}^{\infty}\mathbf{q}^K\sum_{K_1,\dots,K_{\ell}=0\atop K_1+\dots+K_{\ell}=K}\sum_{\vec{Y}_{n}\atop|\vec{Y}_{n}|=K_n}\to \sum_{K=0}^{\infty}\mathbf{q}^K\sum_{\vec{W}\atop|\vec{W}|=K}
\end{equation}
where $\vec{W}$ is a $\ell N$-coloured Young diagram. One can easily convince themselves by example that every possible $\ell N$-tuple with $K$ boxes can be generated by summing over all possible $\ell$ of $N$-tuples $\{\vec{Y}_n\}$ with $K_1,\dots,K_{\ell}$ boxes subject to the constraint $\sum K_n=K$. After all, the set of Young diagrams with $K$ boxes are just counting partitions of $K$ but we can just as well consider a set of $\ell$ sub-Young diagrams counting partitions of $K_1,\dots,K_{\ell}$ and then partitions of those diagrams. Then,
\begin{align}
\Zc^{4d,SU(N)}_{\text{inst},\tilde{A}_{\ell-1}}\left(a_{n,A},m;\mathbf{q}\right)&=\sum_{\{K_n\}}\mathbf{q}^{K}\sum_{\vec{Y}_n\atop|\vec{Y}_n|=K_n}\prod_{n=1}^{\ell}z_{\text{vec}}\left(a_n,\vec{Y}_n\right)z_{\text{bi}}\left(a_{n,A},\vec{Y}_{n};a_{n+1,A},\vec{Y}_{n+1};m\right)\\
&=\sum_{K=0}^{\infty}\mathbf{q}^K\sum_{\vec{W}\atop|\vec{W}|=K} z_{\text{vec}}\left(a_L,\vec{W}\right)z_{\text{bi}}\left(a_L,\vec{W};a_L,\vec{W};m\right)\label{eqn:4drewrite}\\
&=\Zc^{4d,SU(\ell N)}_{\text{inst},\N=2^*}\left(a_L,m;\mathbf{q}\right)
\end{align}
where $z_{\text{vec}}$ is defined in \eqref{eqn:zvec} and
\begin{equation}
z_{\text{bi}}\left(a_n,\vec{Y}_n;a_m,\vec{Y}_m;m\right)\prod_{A,B=1}^{N}\prod_{s\in Y_{n+1,A}}\left(E_{mn,AB}+\frac{\epsilon_+}{2}-m\right)\prod_{s\in Y_{n,B}}\left(E_{nm,AB}+\frac{\epsilon_+}{2}+m\right)
\end{equation}
This is because, for these specific values it is possible to rewrite, for example, 
\begin{equation}
\prod_{n=1}^{\ell}z_{\text{vec}}\left(a_A,\vec{Y}_n\right)=\prod_{A,B=1}^N\frac{1}{\prod_{s_n\in W_{A+nN}}E_{AB}\left(E_{AB}+\epsilon_+\right)}=z_{\text{vec}}\left(a_L,\vec{W}\right)\,,
\end{equation}
where $\vec{W}$ is a $\ell N$-coloured Young diagram given by $\vec{W}=\{\vec{Y}_1,\dots,\vec{Y}_\ell\}=\{Y_{1,1},\dots,Y_{\ell,N}\}$. A similar rewriting may be performed for the bifundamental contribution $z_{\text{bif}}$. Note, for this factorisation to work it is imperative that $\mathbf{q}_A=\mathbf{q}$. Summing over the Young diagrams one arrives at \eqref{eqn:4drewrite}. Note that this may also be seen from the refined topological vertex formalism \cite{Bastian:2017ing}. The $\mathcal{N}_{N,\ell}$ are the 5d lifts of the circular $\tilde{A}_{\ell-1}$ quiver theories and may be obtained by compactifying M-theory on a certain class of non-compact Calabi-Yai 3-folds denoted $X_{N,\ell}$. When $\ell N=\ell'N'$ then $X_{N,\ell}\sim X_{N',\ell'}$ can be related by flop transition and thus $\Zc_{\ell,N}^{\text{top}}(\omega)=\Zc_{\ell',N'}^{\text{top}}(\omega')$.

\subsection{From Class $\mathcal{S}$  to Class $\mathcal{S}_k$ instantons at the orbifold point}
\label{sec:OrbifoldPoint}
\quad
In this section we show that we can write our partition function in a similar form as predicted in \cite{Mitev:2017jqj}. We set $\ell=2$ for simplicity and take a decoupling limit such that we open up the $\tilde{A}_{1}\times \tilde{A}_{k-1}$ in the `$\ell$'-direction such that our Class $\mathcal{S}_k$ theories are given by orbifolds of $\N=2$ SCQCD. This limit is obtained by setting, say, $\mathbf{q}_{4d,1i}=\mathbf{q}_{i}$ and $\mathbf{q}_{4d,2i}=0$. The resulting Class $\mathcal{S}_k$ theories were denoted as SCQCD$_k$ in \cite{Mitev:2017jqj}. If we further go to the `orbifold point' (with respect to the $\mathbb{Z}_k$ orbifold), we set all couplings equal $\mathbf{q}_{i}:=\mathbf{q}/k$ for all $i=1,\dots,k$
\begin{equation}
\Zc^{N}_{\text{inst},\text{$\mathcal{N}=1$ SCQCD$_k$}}(a_{ni,A})=\sum_{K\geq0}\mathbf{q}^{K}\sum_{K_1+\dots+K_k=K}\Zc^{4d,N,2,k}_{\{K_{i}\}}=\Zc^{kN}_{\text{inst},\text{$\mathcal{N}=2$ SCQCD}}(\tilde{a}_{n,\mathcal{A}})\,.
\end{equation}
Furthermore, at the orbifold point the instanton partition function for $\N=1$ $SU(N)^k$ SCQCD$_k$ may be obtained from the orbifold mother $\N=2$ SCQCD $SU(kN)$ theory by making the replacement 
\begin{equation}
a_{\mathcal{A}}\to\tilde{a}_{\mathcal{A}} = a_{A} e^{2\pi\iu j/k}  \,,\quad \mathcal{A}=jA\,,
\end{equation}
where we have identified $\tilde{a}$ with the $\mathbb{Z}_k$ orbifold projection of the vev's of the $SU(kN)$ vector multiplet scalar $\Phi$ of the mother theory.

\section{Conclusions}
\label{sec:Conclusions}

In this paper we computed the instanton partition function of 4d $\mathcal{N}=1$ theories in class $\mathcal{S}_k$ and a 5d and 6d uplift of them, which correspond to 5d $\mathcal{N}=1$ and 6d $(1,0)$ theories in the presence of a half-BPS defect.
We further observed that  class $\mathcal{S}_k$  instanton partition functions can be obtained from the 4d $\mathcal{N}=2$ theories in class $\mathcal{S}$ and their  5d and 6d uplifts: the 5d $\mathcal{N}=1$  necklace quiver $\mathcal{N}_{N,\ell}$ and the 6d $(1,0)$ SCFT $\mathcal{T}^N_{\ell}$ (without the defect) via imposing the `orbifold condition' on the Coulomb moduli and mass parameters as
\begin{equation}
\Zc_{\inst}^{\mathcal{S}_k
, SU(N)}\left(a_{A}\right)=\Zc_{\inst}^{ \mathcal{S},SU(kN) } \left(a_{\mathcal{A}}\right) 
\quad \mbox{with} \quad a_{\mathcal{A}} \rightarrow a_{A} e^{2\pi\iu j/k} 
\end{equation}
with $\mathcal{A} = jA$ being an SU$(kN)$ fundamental index, $A=1,\dots , N$ an SU$(N)$ index and $j=1,\dots , k$ counting the number of the mirror images. 
It is worthwhile to remark that this type of property also holds in the case of $\Z_{\ell}$ orbifold daughters of $SU(\ell N)$ $\N=4$ SYM theory, which are the circular $\N=2$ quivers with gauge group $SU(N)^{\ell}$.
The partition function of mass deformed $\N=4$ SYM, after imposing the orbifold condition on the Coulomb moduli and mass parameters as above, gives the instanton partition function of the circular $\N=2$ quiver
at the orbifold point,
\begin{equation}
\Zc^{\N=2 \,SU(N)^\ell}_{\inst}(a_A)  = \Zc^{\N=2^* \, SU(\ell N)}_{\inst}  (a_L)
 \quad \mbox{with} \quad  a_L=a_{nA}
\end{equation}
where $L=nA=1,\dots,\ell N$, $A=1,\dots , N$ and $n=1,\dots\ell$. This fact was also recently observed in \cite{Bastian:2017ing}. 
This type of simplicity for theories obtained via orbifold constuctions has been long anticipated \cite{Hollowood:1999bm,Fucito:2001ha,Argurio:2007vqa}\footnote{See also \cite{Fucito:2005wc} for similar simplicity for $\mathcal{N}=1^*$.}.

It is  important to stress that our result for the class $\mathcal{S}_k$ instanton partition functions match with the prediction of \cite{Mitev:2017jqj} coming from a calculation of a completely different type.
In \cite{Mitev:2017jqj} based on the anticipation of an AGT type correspondence for theories in class $\mathcal{S}_k$, and the comparison of the spectral curves of theories in  class $\mathcal{S}_k$ with 2d CFT blocks, 
the 2d CFT symmetry algebra and its representations that should underlie AGT$_k$ were identified. 
 These conformal blocks led to  a prediction for the instanton partition functions of the 4d $\mathcal{N}=1$ SCFTs of class $\mathcal{S}_k$ which we precisely reproduce here.
Further work in this direction is definitely worthwhile.

\begin{figure}
\centering
  \begin{tikzpicture}[square/.style={regular polygon,regular polygon sides=4},thick,inner sep=0.1em,scale=1]
  
    \node (G11) at (0,0)[circle,draw,minimum size=1cm]{$N$};
    \node (G21) at (2,0) [circle,draw,minimum size=1cm]{$N$};
    \node (G12) at (0,-2)[circle,draw,minimum size=1cm]{$N$};
    \node (G22) at (2,-2) [circle,draw,minimum size=1cm]{$N$};
    
    \draw [->,color=blue] (G11.280) to (G12.80);
    \draw [<-,color=blue] (G11.260) to (G12.100);
    \draw [->,color=blue] (G21.280) to (G22.80);
    \draw [<-,color=blue] (G21.260) to (G22.100);
    
    \draw [->,color=green] (G11.10) to (G21.170);
    \draw [->,color=green] (G12.10) to (G22.170);
    \draw [<-,color=green] (G11.-10) to (G21.190);
    \draw [<-,color=green] (G12.-10) to (G22.190);
     	
	\draw [<-,color=red] (G11.-55) to (G22.145);
 	\draw [->,color=red] (G11.-35) to (G22.125);
 	\draw [<-,color=red] (G21.-145) to (G12.55);
 	\draw [->,color=red] (G21.-125) to (G12.35);
 	
 	\node at (0,0.75) {$\mathbf{q}_{4d,11}$};
 	\node at (2,0.75) {$\mathbf{q}_{4d,21}$};
 	\node at (2,-2.75) {$\mathbf{q}_{4d,22}$};
 	\node at (0,-2.75) {$\mathbf{q}_{4d,12}$};
 	
 	\draw [->] (3,-1) -- (4,-1) node[midway,above]{$\mathbf{q}_{4d,2i}=0$};
 	\node (G11) at (5,0)[circle,draw,minimum size=1cm]{$N$};
    \node (G21) at (7,0) [square,draw,minimum size=1.4cm]{$2N$};
    \node (G12) at (5,-2)[circle,draw,minimum size=1cm]{$N$};
    \node (G22) at (7,-2) [square,draw,minimum size=1.4cm]{$2N$};
    
    \draw [->,color=blue] (G11.280) to (G12.80);
    \draw [<-,color=blue] (G11.260) to (G12.100);
    
    \draw [->] (G11.10) to (G21.170);
    \draw [->] (G12.10) to (G22.170);
    \draw [<-] (G11.-10) to (G21.190);
    \draw [<-] (G12.-10) to (G22.190);
    
    \node at (5,0.75) {$\mathbf{q}_{4d,11}$};
 	\node at (5,-2.75) {$\mathbf{q}_{4d,12}$};
 	
 	\draw [->] (8,-1) -- (9,-1) node[midway,above]{$\mathbf{q}_{4d,12}=0$};
 	
 	\node (G11) at (10,-1)[circle,draw,minimum size=1cm]{$N$};
    \node (G21) at (12,-1) [square,draw,minimum size=1.4cm]{$3N$};

    \draw [->] (G11.10) to (G21.170);
    \draw [<-] (G11.-10) to (G21.190);    
    \node at (10,-0.25) {$\mathbf{q}_{4d,11}$};     	
  \end{tikzpicture}
  \caption{\normalfont Left: \it $\tilde{A}_1\times\tilde{A}_1$ quiver. \normalfont Middle: \it The quiver may opened up in the $\ell$ direction by taking the decoupling limit $\mathbf{q}_{4d,2i}=0$. The resulting theory is SCQCD$_2$. \normalfont Right: \it  Taking a further decoupling limit $\mathbf{q}_{4d,n2}=0$ yields $\N=1$ SQCD with $N_f=3N$ theory.}
    \label{fig:SQCDlimit}
\end{figure}
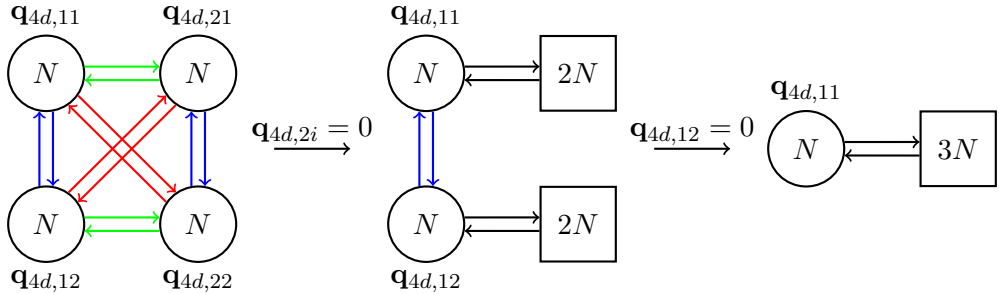

$\mathcal{N}=1$ SQCD with $N_f=3N$ can be obtained from class $\mathcal{S}_k$, from the $\mathbb{Z}_2 \times \mathbb{Z}_2$ theory depicted in Figure \ref{fig:SQCDlimit} in the limit where three of the coupling constants go to zero as shown in the figure.
 It would be very interesting to learn how to isolate the instantons of the $\mathcal{N}=1$ SQCD with $N_f=3N$.  This is one of the most important long term goals of this endeavour. 
 Naively going away from the orbifold point can be obtained by allowing the $\mathbf{q}$ `instanton fugacities' in front of the $K$ D1 brane partition functions to differ.
However, it is not any more clear what the precise  physical interpretation of the $a$ parameters is.
We are currently tying to understand this point \cite{Tom}, by backing up, deriving and studying the SW curves away from the orbifold point, following the work of \cite{Intriligator:1994sm}. 

What is more, we would like to bring to the attention of the reader the fact that the $\mathcal{N}=1$ instanton partition function we derived is a product of an orbifolded vector and a bifundamental hyper multiplet contributions
\begin{equation}
\Zc_{\inst} = \prod_{\text{quiver}}  z^{\text{orb}}_{\text{vec}} \,  z^{\text{orb}}_{\text{bif}}
\end{equation}
directly arising from their $\mathcal{N}=2$ mother theory construction.
An important question is if the instanton partition function  may be further reorganised, for the $\mathcal{S}_k$ theory, as
\begin{equation}
\Zc_{\inst} \stackrel{?}{=} \prod_{\text{quiver}}  z^{\mathcal{N}=1}_{\text{vector}} \, z^{\mathcal{N}=1}_{\text{chiral}} \, .
\end{equation}
At this stage it is unclear to us if this is even possible, however we believe that carefully studying the SW curves away from the orbifold point  \cite{Tom} will be illuminating.

A comment concerning our strategy is in order.
For the computation of the instanton partition function we have decided, instead of computing the matrix model path integral of the 0d theory that lives on the D$(-1)$ branes, to compute the 2d SCI of the gauge theory which lives on the $K$ D1 branes for the following reasons.
First of all, the SCI computation is very well studied/understood and is rather more tractable than directly localising the matrix model. Secondly,
 along the way to the instanton partition function of the 4d theories of interest we also computed the partition function of self-dual strings of a certain set of 6d $(1,0)$ theories in the presence of half-BPS surface operators, which is a very interesting result in its own right. 
 It would be very interesting to attempt localisation and compute the 2d partition function that we have computed, in analogy with the localisation computation of the Lens space index \cite{Benini:2011nc,Alday:2013rs}.  Another very interesting alternative to our strategy is the Origami approach of Nekrasov \cite{Nekrasov:2016ydq}.

Our strategy may be applied to several other interesting cases such as computing partition functions of $(1,0)$ theories in the presence of a surface operator lying along $\mathbb{C}\subset\mathbb{C}^2_{\epsilon_1\epsilon_2}$. 
Some results along these lines already exist in the literature \cite{Kanno:2011fw,Pan:2016fbl,LeFloch:2017lbt,Gorsky:2017hro,Bawane:2017gjf}. 
The partition function of the $(1,0)$ theory on $T^2$ in the presence of certain half-BPS surface operator should be related to a certian orbifold of the M-string ellipic genus \cite{Haghighat:2013gba,Haghighat:2013tka}, this is currently being explored in \cite{Tom2}.
Finally, one could also try to compute partition functions in class $\mathcal{S}_k$ in the presence of defects via combining the two orbifold constructions.

In an orthogonal direction, and in connection with \cite{Mitev:2017jqj}, it would be instructive to try to repeat  the method of \cite{Cordova:2016cmu}, who, starting from the $(2,0)$ theory in 6d, were able to obtain a direct derivation of the AGT correspondence,
for the $\mathcal{N}=1$ theories of class $\mathcal{S}_k$.

Finally, using our results and taking the large $N$ limit, one could learn about the gravity dual of $\mathcal{N}=1$ theories in class $\mathcal{S}_k$ following the work of \cite{Dorey:1999pd,Hollowood:1999bm}.

\section{Acknowledgements}
It is a great pleasure to thank Giulio Bonelli, Jan Peter Carstensen,  Troy Figiel,  Bruno Le Floch,     Marialuisa Frau, Amihay Hanany, 
Alberto Lerda,  Wolfger Peelaers, Shlomo  Razamat, Martin Rocek, Alessandro Tanzini 
and Futoshi Yagi for useful discussions.
Our work is supported by the German Research Foundation (DFG) via the Emmy
Noether program ``Exact results in Gauge theories''.
Finally, we thank the organizers of the Pollica Summer
Workshop 2017, where some of this work was done,  for hospitality.  
During the workshop we were partly supported by the ERC STG grant 306260.

\appendix

\section{The supersymmetric index computation}\label{App:EllipticGenus}

\subsection{Single letter indices}
As the Witten index is independent of coupling constants we may compute the index in the free field $g\to0$ limit. 
To compute the index we list the gauge covariant field content with $\delta=H_+-\frac{1}{2}\mathfrak{R}=0$. Only the `letters' with $\delta=0$ contribute to the index and their
quantum numbers are listed in Tables \ref{tab:Vletters}, \ref{tab:Hletters} and \ref{tab:Uletters}. We denote also the decomposition of $\N=(4,4)$ multiplets into $\N=(0,2)$ multiplets.
Since the $\N=(0,2)$ field strength multiplet is not conformal extra care must be taken to take the free field limit. In two dimensions the field strength multiplet $\Upsilon$ is nothing but a Fermi multiplet with auxillary $D-\iu F_{+-}$ . The $R$-charge of $\Upsilon$ is fixed to unity everywhere along the flow, i.e. $\mathfrak{R}\left[\Upsilon\right]=\overline{J}_0\left[\Upsilon\right]=1$ . Therefore the index of the off-diagonal vector multiplet with Cartan zero modes should be equal to that of a off-diagonal Fermi multiplet of $R$-charge $\mathfrak{R}_{\text{IR}}=\mathfrak{R}_{\text{UV}}=1$:
\begin{equation}
\Zc_{\text{vec}}(y_I\neq y_J,q)=\Delta(y)^{-1}\Zc_{\text{Fermi}}(y_I\neq y_J,q)=\PE\left[-\frac{2q}{1-q}\sum_{I\neq J}\frac{y_I}{y_J}\right]
\end{equation}
where $\Delta(y)$ is the Vandermonde determinant accounting for the Cartan zero modes.
\begin{table}
\centering
 \begin{tabular}{|c|c|c|c|c|c|c|} 
 \hline
$\N=(0,2)$ & Letter & $\overline{L}_0$& $\overline{J}_0$ & $\gamma_{\ell}$&$\gamma_k$ & Index\\\hline
\hline
\multirow{ 3}{*}{$Y,Y^{\dagger}$}&$Y^{2\dot2}$ &$0$ & $0$& $+1$ &$0$& $wz$\\\cline{2-7}
&$Y^{1\dot1}$ &$0$ & $0$& $-1$ & $0$ & $w^{-1}z^{-1}$\\\cline{2-7}
&$\overline{\lambda}_{+\frac{1}{2}}^{\dot11}$ &$1/2$ & $+1$ & $-1$ & $0$ & $-w^{-1}z^{-1}$\\\hline
\multirow{ 3}{*}{$\tilde{Y}^{\dagger},\tilde{Y}$}&$Y^{1\dot2}$ & $0$ & $0$& $-1$& $-1$ & $w^{-1}z$\\\cline{2-7}
&$Y^{2\dot1}$ &$0$ & $0$& $+1$ & $+1$ & $wz^{-1}$\\\cline{2-7}
&$\overline{\lambda}_{+\frac{1}{2}}^{\dot12}$ &$1/2$ & $+1$ & $+1$& $+1$&$-wz^{-1}$\\\hline
\multirow{ 2}{*}{$\overline{\xi},\overline{\xi}^{\dagger}$} &$\overline{\xi}_{-\frac{1}{2}}^{\dot1\dot1}$ &$0$ & $0$ & $0$ & $0$ & $-qz^{-2}$\\\cline{2-7}
&$\overline{\xi}_{-\frac{1}{2}}^{\dot2\dot2}$ &$0$ & $0$ & $0$ & $-1$ & $-z^2$\\\hline
\multirow{ 2}{*}{$\Upsilon,\Upsilon^{\dagger}$}&$\overline{\xi}_{-\frac{1}{2}}^{\dot2\dot1}$ &$-1/2$ & $-1$ & $0$ & $-1$ & $-q$\\\cline{2-7}
&$\overline{\xi}_{-\frac{1}{2}}^{\dot1\dot2}$ &$1/2$ & $+1$ & $0$ &$-1$& $-q$\\\hline
\hline
\multirow{ 2}{*}{}&$\partial_-\overline{\lambda}_{+\frac{1}{2}}^{\dot11}$ &$1/2$ & $+1$ & $-1$& $-1$&$qw^{-1}z^{-1}$\\\cline{2-7}
&$\partial_-\overline{\lambda}_{+\frac{1}{2}}^{\dot12}$ &$1/2$ & $+1$ & $+1$& $0$& $qwz^{-1}$\\\hline
\hline
&$\partial_{-}$ &$0$ &$0$& $0$ & $-1$ & $q$\\\hline
\end{tabular}
\caption{\it Gauge covariant field content with $\delta=0$ of the $\N=(4,4)$ vector multiplet $V$.}
 \label{tab:Vletters}
\end{table}
\begin{table}
\centering
 \begin{tabular}{|c|c|c|c|c|c|c|} 
 \hline
$\N=(0,2)$ &Letter & $\overline{L}_0$& $\overline{J}_0$ & $\gamma_{\ell}$ &$\gamma_k$& Index\\\hline
\hline
\multirow{ 3}{*}{$X,X^{\dagger}$}&$X^{1\dot1}$ &$0$ & $0$& $0$ & $0$& $q^{\frac{1}{2}}v^{-1}z^{-1}$\\\cline{2-7}
&$X^{2\dot2}$ &$0$ & $0$& $0$&$0$ & $q^{-\frac{1}{2}}vz$\\\cline{2-7}
&$\xi_{+\frac{1}{2}}^{2\dot2}$ &$1/2$ & $+1$ & $0$& $0$ &$-q^{-\frac{1}{2}}vz$\\\hline
\multirow{ 3}{*}{$\tilde{X}^{\dagger},\tilde{X}$}&$X^{2\dot1}$ &$0$ & $0$& $0$&$0$ & $q^{\frac{1}{2}}vz^{-1}$\\\cline{2-7}
&$X^{1\dot2}$ &$0$ & $0$& $0$ & $0$&$q^{-\frac{1}{2}}v^{-1}z$\\\cline{2-7}
&$\xi_{+\frac{1}{2}}^{1\dot2}$ &$1/2$ & $+1$ & $0$&$0$& $-q^{-\frac{1}{2}}v^{-1}z$\\\hline
\multirow{ 2}{*}{$\lambda,\lambda^{\dagger}$} &$\lambda_{-\frac{1}{2}}^{11}$ &$0$ & $0$ & $-1$ &$-1$& $-q^{\frac{1}{2}}v^{-1}w^{-1}$\\\cline{2-7}
&$\lambda_{-\frac{1}{2}}^{22}$ &$0$ & $0$ & $+1$ & $0$ &$-q^{\frac{1}{2}}vw$\\\hline
\multirow{ 2}{*}{$\tilde{\lambda}^{\dagger},\tilde{\lambda}$}&$\lambda_{-\frac{1}{2}}^{12}$ &$0$ & $0$ & $+1$ & $0$&$-q^{\frac{1}{2}}v^{-1}w$\\\cline{2-7}
&$\lambda_{-\frac{1}{2}}^{21}$ &$0$ & $0$ & $-1$ & $0$ & $-q^{\frac{1}{2}}vw^{-1}$\\\hline
\hline
\multirow{ 2}{*}{}&$\partial_-\xi_{+\frac{1}{2}}^{2\dot2}$ &$1/2$ & $+1$ & $0$& $-1$&$q^{\frac{1}{2}}vz$\\\cline{2-7}
&$\partial_-\xi_{+\frac{1}{2}}^{1\dot2}$ &$1/2$ & $+1$ & $0$&$-1$& $q^{\frac{1}{2}}v^{-1}z$\\\hline
\hline
&$\partial_{-}$ &$0$ &$0$& $0$ &$-1$& $q$\\\hline
\end{tabular}
\caption{\it Gauge covariant field content with $\delta=0$ of the $\N=(4,4)$ hypermultiplet $H$.}
 \label{tab:Hletters}
\end{table}
\begin{table}
\centering
 \begin{tabular}{|c|c|c|c|c|c|c|} 
 \hline
$\N=(0,2)$&Letter & $\overline{L}_0$& $\overline{J}_0$ & $\gamma_{\ell}$&$\gamma_k$ & Index\\\hline
\hline
\multirow{ 3}{*}{$\phi,\phi^{\dagger}$}&$\phi^{\dot1}$ &$0$ & $0$& $0$&$0$ & $q^{\frac{1}{2}}z^{-1}$\\\cline{2-7}
&$\phi^{\dagger}_{\dot1}$ &$0$ & $0$& $0$ & $0$&$q^{-\frac{1}{2}}z$\\\cline{2-7}
&$\chi^{\dagger}_{+\frac{1}{2}\dot1}$ &$1/2$ & $+1$ & $0$ &$0$& $-q^{-\frac{1}{2}}z$\\\hline
\multirow{ 3}{*}{$\tilde{\phi}^{\dagger},\phi$}&$\phi^{\dagger}_{\dot2}$ &$0$ & $0$& $0$&$0$ & $q^{\frac{1}{2}}z^{-1}$\\\cline{2-7}
&$\phi^{\dot2}$ &$0$ & $0$& $0$ &$0$& $q^{-\frac{1}{2}}z$\\\cline{2-7}
&$\chi_{+\frac{1}{2}}^{\dot2}$ &$1/2$ & $+1$ & $0$& $0$&$-q^{-\frac{1}{2}}z$\\\hline
\multirow{ 2}{*}{$\psi,\psi^{\dagger}$}&$\psi_{-\frac{1}{2}}^{1}$ &$0$ & $0$ & $-1$ &$-1$& $-q^{\frac{1}{2}}w^{-1}$\\\cline{2-7}
&$\psi^{\dagger}_{-\frac{1}{2}1}$ &$0$ & $0$ & $1$ &$0$& $-q^{\frac{1}{2}}w$\\\hline
\multirow{ 2}{*}{$\tilde{\psi}^{\dagger},\tilde{\psi}$}&$-\psi^{\dagger}_{-\frac{1}{2}2}$ &$0$ & $0$ & $-1$ &$-1$& $-q^{\frac{1}{2}}w^{-1}$\\\cline{2-7}
&$\psi_{-\frac{1}{2}}^{2}$ &$0$ & $0$ & $1$ &$0$& $-q^{\frac{1}{2}}w$\\\hline
\hline
\multirow{ 2}{*}{}&$\partial_-\chi^{\dagger}_{+\frac{1}{2}\dot1}$ &$1/2$ & $+1$ & $0$&$-1$& $q^{\frac{1}{2}}z$\\\cline{2-7}
&$\partial_-\chi_{+\frac{1}{2}}^{\dot2}$ &$1/2$ & $+1$ & $0$&$-1$& $q^{\frac{1}{2}}z$\\\hline
\hline
&$\partial_{-}$ &$0$ &$0$& $0$&$-1$ & $q$\\\hline
\end{tabular}
\caption{\it Gauge covariant field content with $\delta=0$ of the $\N=(4,4)$ hypermultiplet $U$.}
 \label{tab:Uletters}
\end{table}
The single letter indices we given in \eqref{eqn:letterfV}, \eqref{eqn:letterfH} and \eqref{eqn:letterfV}. We again list them here
\begin{align}
&i_V(q,w,z,y_I)=\left[\frac{\left(w+w^{-1}\right)\left(z+qz^{-1}\right)-qz^{-2}-z^2-2q}{1-q}\right]\sum_{I,J=1}^Ky_Iy_J^{-1}\,,\\
&i_H(q,v,w,z,y_I)=\left[\frac{q^{\frac{1}{2}}\left(v+v^{-1}\right)\left(z+z^{-1}-w^{-1}-w\right)}{1-q}\right]\sum_{I,J=1}^Ky_Iy_J^{-1}\,,\\
&i_U(q,w,z,x_A,y_I)=\left[\frac{q^{\frac{1}{2}}\left(z+z^{-1}-w^{-1}-w\right)}{1-q}\right]\sum_{I=1}^K\sum_{A=1}^N\left(y_Ix_A^{-1}+y_I^{-1}x_A\right)\,.
\end{align}
Finally, we also list the Casimir energy:
\begin{equation}
E_{\text{Casimir}}=\substack{\text{Finite}\\q\to1}\left[\sum_{\mathcal{M}}\frac{\partial i_{\mathcal{ M}}}{\partial\log q}\right]=\frac{\beta^2_5}{\iu\pi}2NK\left(\frac{\epsilon_+}{2}+m\right)\left(\frac{\iu\pi}{\beta_5}+\frac{\epsilon_+}{2}-m\right)\label{eqn:Casimir1} \,.
\end{equation}
\subsection{Orbifolded single letter indices}
The single letters for the $\Gamma$ projected multiplets is given by enumerating all letters in Tables \ref{tab:Vletters}, \ref{tab:Hletters} and \ref{tab:Uletters} while also inserting fugacities for the $\Gamma$ action embedded in the global and gauge symmetries. Recall that 
\begin{equation}
\gamma_{\ell}:=2J_L^R=J_{710}-J_{89}\,,\quad \gamma_k:=J_{56}+J_L^R-J_R^R=J_{56}-J_{89}\,.
\end{equation}
The projected single letters are thus given by
\begin{equation}
\begin{aligned}
i_{\Gamma V}(q,w,z,y_{ni,I})=&\frac{1}{\ell k}\sum_{\varepsilon\in\Z_{\ell}\atop{\varepsilon_k\in\Z_k}}\left[\left(\varepsilon_{\ell}w+\varepsilon^{-1}_{\ell}\varepsilon_k^{-1}w^{-1}\right)\left(z+qz^{-1}\right)-qz^{-2}-\varepsilon_k^{-1}z^2-\varepsilon_k^{-1}2q\right]\\
&\times\sum_{r\geq0}q^r\varepsilon_k^{-r}\sum_{n,i}^{\ell}\sum_{m,j}^k\sum_{I=1}^{K_{ni}}\sum_{I=1}^{K_{mj}}\varepsilon_{\ell}^{n-m}\varepsilon_k^{i-j}y_{ni,I}y_{mj,J}^{-1}\,,
\end{aligned}
\end{equation}
\begin{equation}
\begin{aligned}
i_{\Gamma H}(q,v,w,z,y_{ni,I})=&\frac{1}{\ell k}\sum_{\varepsilon\in\Z_{\ell}\atop{\varepsilon_k\in\Z_k}}\left[q^{\frac{1}{2}}\left(v+v^{-1}\right)\left(z^{-1}+\varepsilon_k^{-1}z-\varepsilon_{\ell}^{-1}\varepsilon_k^{-1}w^{-1}-\varepsilon_{\ell}w\right)\right]\\
&\times\sum_{r\geq0}q^r\varepsilon_k^{-r}\sum_{n,m}^{\ell}\sum_{i,i}^k\sum_{I=1}^{K_{ni}}\sum_{I=1}^{K_{mj}}\varepsilon_{\ell}^{n-m}\varepsilon_k^{i-j}y_{ni,I}y_{jm,J}^{-1}\,,
\end{aligned}
\end{equation}
\begin{equation}
\begin{aligned}
i_{\Gamma U}(q,w,z,x_{ni,A},y_{ni,I})=&\frac{1}{\ell k}\sum_{\varepsilon\in\Z_{\ell}\atop{\varepsilon_k\in\Z_k}}\left[q^{\frac{1}{2}}\left(\varepsilon_k^{-1}z+z^{-1}-\varepsilon_{\ell}^{-1}\varepsilon_k^{-1}w^{-1}-\varepsilon_{\ell}w\right)\right]\sum_{n\geq0}q^n\varepsilon_k^{-n}\\
&\times\sum_{n,m}^{\ell}\sum_{i,j}^k\sum_{A=1}^{N}\varepsilon_{\ell}^{n-m}\varepsilon_k^{i-j}\left(\sum_{I=1}^{K_{ni}}y_{ni,I}x_{mj,A}^{-1}+\sum_{I=1}^{K_{mj}}y_{mj,I}^{-1}x_{ni,A}\right)\,.
\end{aligned}
\end{equation}
We now detail how to evaluate the sums over conformal descendants and over the orbifold group.
Firstly, to evaluate the sums over conformal descendants we write to $r:=L_{ij}+\tilde{r}k\geq0$ with $L_{ij}$ defined in \eqref{eqn:Ljv}. This enables one to rewrite, for any fixed value $1\leq j\leq k$, to split the sum 
\begin{equation}
\sum_{r\geq0}q^r\varepsilon_k^{-r}=\sum_{i=1}^kq^{L_{ij}}\varepsilon_k^{-L_{ij}}\sum_{\tilde{r}\geq0}q^{\tilde{r}k}=\sum_{i=1}^k\frac{q^{L_{ij}}\varepsilon_k^{-L_{ij}}}{1-q^k}\,, 
\end{equation}
recall that $\varepsilon_k^k=1$.
After this rewriting the sums over both $\Z_{\ell},\Z_{k}$ may be simply carried out and is essentially equivalent to demanding that the exponents of $\varepsilon_{\ell}$, $\varepsilon_k$ vanish modulo $\ell$, $k$ in each term. Hence we have after, rearranging and applying the identity \eqref{eqn:usefulid},
\begin{equation}
\begin{aligned}
i_{\Gamma V}&\left(q,w,z,y_{ni,I}\right)=\\
&\sum_{n=1}^{\ell}\sum_{i,j=1}^k\frac{1}{1-q^k}\left[-\sum_{I=1}^{K_{ni}}\sum_{J=1}^{K_{nj}}\left(z^{-2}q^{L_{ij}+1}y_{ni,I}y_{nj,J}^{-1}+z^2q^{k-L_{ij}-1}y_{ni,I}^{-1}y_{nj,J}\right)\right.\\
&+\sum_{I=1}^{K_{ni}}\sum_{J=1}^{K_{(n+1)j}}\left(wq^{L_{ij}}y_{ni,I}y_{(n+1)j,J}^{-1}+w^{-1}q^{k-L_{ij}-1}y_{ni,I}^{-1}y_{(n+1)j,J}\right)\left(z+qz^{-1}\right)\\
&\left.-\sum_{I=1}^{K_{ni}}\sum_{J=1}^{K_{nj}}\left(q^{L_{ij}}y_{ni,I}y_{nj,J}^{-1}+\left(q^{k-L_{ij}}-(1-q^k)\delta_{L_{ni},0}\right)y_{ni,I}^{-1}y_{nj,J}\right)\right],
\end{aligned}\label{eqn:orbletterfV}
\end{equation}
\begin{equation}
\begin{aligned}
i_{\Gamma H}&\left(q,v,w,z,y_{ni,I}\right)=\\
&\sum_{n=1}^{\ell}\sum_{i,j=1}^k\frac{q^{\frac{1}{2}}\left(v+v^{-1}\right)}{1-q^k}\left[\sum_{I=1}^{K_{ni}}\sum_{J=1}^{K_{nj}}\left(z^{-1}q^{L_{ij}}y_{ni,I}y_{nj,J}^{-1}+zq^{k-L_{ij}-1}y_{ni,I}^{-1}y_{nj,J}\right)\right.\\
&\left.-\sum_{I=1}^{K_{ni}}\sum_{J=1}^{K_{(n+1)j}}\left(wq^{L_{ij}}y_{ni,I}y_{(n+1)j,J}^{-1}+w^{-1}q^{k-L_{ij}-1}y_{ni,I}^{-1}y_{(n+1)j,J}\right)\right],
\end{aligned}\label{eqn:orbletterfH}
\end{equation}
\begin{equation}
\begin{aligned}
i_{\Gamma U}&\left(q,w,z,x_{ni,A},y_{ni,I}\right)=\sum_{n=1}^{\ell}\sum_{i,j=1}^k\sum_{A=1}^N\frac{q^{\frac{1}{2}}}{1-q^k}\\
&\left[z^{-1}q^{L_{ij}}\left(\sum_{I=1}^{K_{ni}}y_{ni,I}x_{nj,A}^{-1}+\sum_{I=1}^{K_{nj}}y_{nj,I}^{-1}x_{ni,A}\right)+zq^{k-L_{ij}-1}\left(\sum_{I=1}^{K_{ni}}y_{ni,I}^{-1}x_{nj,A}+\sum_{I=1}^{K_{nj}}y_{nj,I}x_{ni,A}^{-1}\right)\right.\\
&-wq^{L_{ij}}\left(\sum_{I=1}^{K_{ni}}y_{ni,I}x_{(n+1)j,A}^{-1}+\sum_{I=1}^{K_{(n+1)j}}y_{(n+1)j,I}^{-1}x_{ni,A}\right)\\
&\left.-w^{-1}q^{k-L_{ij}-1}\left(\sum_{I=1}^{K_{ni}}y_{ni,I}^{-1}x_{(n+1)j,A}+\sum_{I=1}^{K_{(n+1)j}}y_{(n+1)j,I}x_{ni,A}^{-1}\right)\right].
\end{aligned}\label{eqn:orbletterfU}
\end{equation}
In this form the plethystics may be easily performed. For the sake of completeness we also list the contribution from the Casimir energy \eqref{eqn:casimir}
\begin{equation}
\begin{aligned}E_{\text{Casimir}}=&\frac{k\beta^2_5}{\iu\pi}\left(2NkK\left(\frac{\epsilon_+}{2}+m\right)\left(\frac{\iu\pi}{\beta_5}+\frac{\epsilon_+}{2}-m\right)\right)\\
&+\frac{k\beta_5^2}{\pi^2}\sum_{n=1}^{\ell}\sum_{\mathcal{A}=1}^{kN}\sum_{\mathcal{I}=1}^{K_n}u_{n,\mathcal{I}}\left(\tilde{a}_{n+1,\mathcal{A}}+\tilde{a}_{n-1,\mathcal{A}}-2\tilde{a}_{n,\mathcal{A}}\right)\\
&+\frac{k\beta_5^2}{\pi^2}\sum_{n=1}^{\ell}\sum_{\mathcal{A}=1}^{kN}K_n\left(2\tilde{a}_{n,\mathcal{A}}^2-\tilde{a}_{n-1,\mathcal{A}}^2-\tilde{a}_{n+1,\mathcal{A}}^2+2m\tilde{a}_{n+1,\mathcal{A}}-2m\tilde{a}_{n-1,\mathcal{A}}\right)\label{eqn:lkCasimir} 
\end{aligned}
\end{equation}
where we also made the gauge transformation and redefinition \eqref{eqn:gaugetrans} and used the definitions \eqref{eqn:lk5dvariables}.

\section{4d \& 5d contour integral representations}

In this appendix we present the contour integral representations for the partition functions for the 5d and 4d theories both in the precense of the orbifold and without. These may be obtained by applying the limit directly to the respective 6d contour integral expression. We follow mostly the prescription presented in \cite{Gadde:2011ia}. We will firstly take the 5d $\beta_6\to0$ $(q\to1)$ limit.

\subsection{5d limit of the unorbifolded contour integral}
\label{5dLimitOfUnorbifolded}

Using the identifications \eqref{eqn:5dvariables} and setting $y_I=q^{\frac{\beta_5u_I}{\iu\pi}}$ we have that
\begin{align}
&\lim_{q\to1}\prod_{I\neq J}\left(1-\frac{y_I}{y_J}\right)\Zc_V=\prod_{I\neq J}\sinh\beta_5\left(u_{IJ}\right)\prod_{I,J=1}^K\frac{\sinh\beta_5\left(u_{IJ}-\epsilon_+\right)}{\sinh\beta_5\left(u_{IJ}-\frac{\epsilon_+}{2}-m\right)\sinh\beta_5\left(u_{IJ}+\frac{\epsilon_+}{2}-m\right)}\,,\\
&\lim_{q\to1}\Zc_H=\prod_{I,J=1}^K\frac{\sinh\beta_5\left(u_{IJ}+\frac{\epsilon_-}{2}+m\right)\sinh\beta_5\left(u_{IJ}-\frac{\epsilon_-}{2}+m\right)}{\sinh\beta_5\left(u_{IJ}+\epsilon_1\right)\sinh\beta_5\left(u_{IJ}+\epsilon_2\right)}\,,\\
&\lim_{q\to1}\Zc_U=\prod_{I=1}^K\prod_{A=1}^N\frac{\sinh\beta_5\left(u_{I}-a_A-m\right)\sinh\beta_5\left(u_{I}-a_A+m\right)}{\sinh\beta_5\left(u_{I}-a_A-\frac{\epsilon_+}{2}\right)\sinh\beta_5\left(u_{I}-a_A+\frac{\epsilon_+}{2}\right)}\,,
\end{align}
where $u_{IJ}:=u_I-u_J$.
By definition 
\begin{equation}
\lim_{q\to1}\Zc^{(0)}(q,v,w,z,x_A,y_I)=1\,.
\end{equation}
Hence, all that remains is to perform the limit on the integration over the maximal torus of $U(K)$:
\begin{equation}
\lim_{q\to1}\oint_{T\left[U\left(K\right)\right]}\prod_{I=1}^K\frac{dy_I}{2\pi\iu y_I}=\lim_{\beta_6\to0}\left(2\tau\right)^K\int^{\frac{\iu\pi}{2\tau}}_{-\frac{\iu\pi}{2\tau}}\prod_{I=1}^K\frac{du_I}{2\pi\iu\beta_5}=\int^{+\infty}_{-\infty}\prod_{I=1}^K\frac{du_I}{2\pi\iu\beta_5}\,.
\end{equation}

Putting all of the above ingredients together we write
\begin{align}\label{eqn:5dpartitionfunction}
&\Zc^{5d,N}_K:=\lim_{q\to1}\Zc^{6d,N}_K\\
&\begin{aligned}
&=\sum_{K\geq0}\frac{1}{K!}\int\prod_{I=1}^K\frac{du_I}{2\pi\iu\beta_5}\prod_{I=1}^K\prod_{A=1}^N\frac{\sinh\beta_5\left(u_{I}-a_A-m\right)\sinh\beta_5\left(u_{I}-a_A+m\right)}{\sinh\beta_5\left(u_{I}-a_A-\frac{\epsilon_+}{2}\right)\sinh\beta_5\left(u_{I}-a_A+\frac{\epsilon_+}{2}\right)}\prod_{I\neq J}\sinh\beta_5\left(u_{IJ}\right)\\
&\times\prod_{I,J=1}^K\frac{\sinh\beta_5\left(u_{IJ}-\epsilon_+\right)\sinh\beta_5\left(u_{IJ}+\frac{\epsilon_-}{2}+m\right)\sinh\beta_5\left(u_{IJ}-\frac{\epsilon_-}{2}+m\right)}{\sinh\beta_5\left(u_{IJ}-\frac{\epsilon_+}{2}-m\right)\sinh\beta_5\left(u_{IJ}+\frac{\epsilon_+}{2}-m\right)\sinh\beta_5\left(u_{IJ}+\epsilon_1\right)\sinh\beta_5\left(u_{IJ}+\epsilon_2\right)}.
\end{aligned}
\end{align}

\subsection{4d limit of the unorbifolded contour integral}
It is then a straightforward exercise to take the 4d limit $\beta_5\to0$. We have
\begin{align}
&\Zc^{4d,N}_K:=\lim_{\beta_5\to0}\Zc^{5d,N}_K\label{eqn:4dpartitionfunction}\\
&\begin{aligned}
&=\sum_{K\geq0}\frac{1}{K!}\int\prod_{I=1}^K\frac{du_I}{2\pi\iu}\prod_{I=1}^K\prod_{A=1}^N\frac{\left(u_{I}-a_A-m\right)\left(u_{I}-a_A+m\right)}{\left(u_{I}-a_A-\frac{\epsilon_+}{2}\right)\left(u_{I}-a_A+\frac{\epsilon_+}{2}\right)}\\
&\times\prod_{I\neq J}u_{IJ}\prod_{I,J=1}^K\frac{\left(u_{IJ}-\epsilon_+\right)\left(u_{IJ}-\frac{\epsilon_-}{2}-m\right)\left(u_{IJ}+\frac{\epsilon_-}{2}-m\right)}{\left(u_{IJ}-m-\frac{\epsilon_+}{2}\right)\left(u_{IJ}+m-\frac{\epsilon_+}{2}\right)\left(u_{IJ}-\epsilon_1\right)\left(u_{IJ}-\epsilon_2\right)}\,.
\end{aligned}
\end{align}

\subsection{5d limit of the orbifolded contour integral}
Taking this limit is largely the same procedure as for the $\ell=k=1$ case however we instead use the slightly different set of variables \eqref{eqn:lk5dvariables}.
We are again interested in the $q\to1$ limit of the partition function \eqref{eqn:lkmnpartfnrewrite}. Setting $y_{i,\mathcal{I}}=q^{\frac{ku_{i\mathcal{I}}}{\iu\pi}}$ we have
\begin{align}
&\lim_{q\to1}\Zc^{6d,N,\ell,k}_{\{K_{ij}\}}:=\Zc^{5d,N,\ell,k}_{\{K_{ij}\}}\label{eqn:lknm5dpartitionfunction}\\
&\begin{aligned}
=&\prod_{i=1}^{\ell}\Bigg[\frac{1}{\prod_{j=1}^kK_{ij}!}\int\prod_{\mathcal{I}=1}^{K_{i}}\frac{du_{i,\mathcal{I}}}{2\pi\iu\beta_5}\prod_{\mathcal{I}\neq \mathcal{J}}\sinh\beta_5\left(u_{i,\mathcal{I}}-u_{i,\mathcal{J}}\right)\\
&\times\prod_{\mathcal{I},\mathcal{J}=1}^{K_{i}}\frac{\sinh\beta_5\left(u_{i,\mathcal{I}}-u_{i,\mathcal{J}}-\epsilon_+\right)}{\sinh\beta_5\left(u_{i,\mathcal{I}}-u_{i,\mathcal{J}}+\epsilon_2\right)\sinh\beta_5\left(u_{i,\mathcal{I}}-u_{i,\mathcal{J}}+\epsilon_1\right)}\\
&\times\prod_{\mathcal{I}=1}^{K_{i}}\prod_{\mathcal{J}=1}^{K_{i+1}}\frac{\sinh\beta_5\left(u_{i,\mathcal{I}}-u_{i+1,\mathcal{J}}+m+\frac{\epsilon_-}{2}\right)\sinh\beta_5\left(u_{i,\mathcal{I}}-u_{i+1,\mathcal{J}}+m-\frac{\epsilon_-}{2}\right)}{\sinh\beta_5\left(u_{i,\mathcal{I}}-u_{i+1,\mathcal{J}}+\frac{\epsilon_+}{2}+m\right)\sinh\beta_5\left(u_{i,\mathcal{I}}-u_{i+1,\mathcal{J}}-\frac{\epsilon_+}{2}+m\right)}\\
&\times\prod_{\mathcal{A}=1}^{kN}\prod_{\mathcal{I}=1}^{K_{i}}\frac{\sinh\beta_5\left(u_{i,\mathcal{I}}-\tilde{a}_{i+1,\mathcal{A}}+m\right)\sinh\beta_5\left(u_{i,\mathcal{I}}-\tilde{a}_{i-1,\mathcal{A}}-m\right)}{\sinh\beta_5\left(u_{i,\mathcal{I}}-\tilde{a}_{i,\mathcal{A}}-\frac{\epsilon_+}{2}\right)\sinh\beta_5\left(u_{i,\mathcal{I}}-\tilde{a}_{i,\mathcal{A}}+\frac{\epsilon_+}{2}\right)}\Bigg]\,.
\end{aligned}
\end{align}

\subsection{4d limit of the orbifolded contour integral}
As before it is straightforward to take the 4d limit $\beta_5\to0$.
\begin{align}
&\Zc^{4d,N,\ell,k}_{\{K_{ij}\}}:=\lim_{\beta_5\to0}\Zc^{5d,N,\ell,k}_{\{K_{ij}\}}\\
&\begin{aligned}
=&\prod_{i=1}^{\ell}\Bigg[\frac{1}{\prod_{j=1}^kK_{ij}!}\int\prod_{\mathcal{I}=1}^{K_{i}}\frac{du_{i,\mathcal{I}}}{2\pi\iu}\prod_{\mathcal{I}\neq \mathcal{J}}\left(u_{i,\mathcal{I}}-u_{i,\mathcal{J}}\right)\prod_{\mathcal{I},\mathcal{J}=1}^{K_{i}}\frac{\left(u_{i,\mathcal{I}}-u_{i,\mathcal{J}}-\epsilon_+\right)}{\left(u_{i,\mathcal{I}}-u_{i,\mathcal{J}}+\epsilon_2\right)\left(u_{i,\mathcal{I}}-u_{i,\mathcal{J}}+\epsilon_1\right)}\\
&\times\prod_{\mathcal{I}=1}^{K_{i}}\prod_{\mathcal{J}=1}^{K_{i+1}}\frac{\left(u_{i,\mathcal{I}}-u_{i+1,\mathcal{J}}+m+\frac{\epsilon_-}{2}\right)\left(u_{i,\mathcal{I}}-u_{i+1,\mathcal{J}}+m-\frac{\epsilon_-}{2}\right)}{\left(u_{i,\mathcal{I}}-u_{i+1,\mathcal{J}}+\frac{\epsilon_+}{2}+m\right)\left(u_{i,\mathcal{I}}-u_{i+1,\mathcal{J}}-\frac{\epsilon_+}{2}+m\right)}\\
&\times\prod_{\mathcal{A}=1}^{kN}\prod_{\mathcal{I}=1}^{K_{i}}\frac{\left(u_{i,\mathcal{I}}-\tilde{a}_{i+1,\mathcal{A}}+m\right)\left(u_{i,\mathcal{I}}-\tilde{a}_{i-1,\mathcal{A}}-m\right)}{\left(u_{i,\mathcal{I}}-\tilde{a}_{i,\mathcal{A}}-\frac{\epsilon_+}{2}\right)\left(u_{i,\mathcal{I}}-\tilde{a}_{i,\mathcal{A}}+\frac{\epsilon_+}{2}\right)}\Bigg]\,.
\end{aligned}
\end{align}


\providecommand{\href}[2]{#2}\begingroup\raggedright\endgroup

\end{document}